\documentclass[twocolumn,prb,amsmath,amssymb,longbibliography,floatfix,superscriptaddress]{revtex4-2}
\usepackage{amsmath}
\usepackage{amssymb}
\usepackage{color}
\usepackage{physics}
\usepackage{graphicx}
\usepackage{tabularx}
\usepackage{dcolumn} 
\usepackage{bm}      
\usepackage{array}   
\usepackage{amsfonts}
\usepackage{url}
\usepackage[bookmarks=false,colorlinks,citecolor=red]{hyperref}
\usepackage{mathtools}
\usepackage{subcaption}
\usepackage{float}

\usepackage[color=green!60]{todonotes}
\usepackage[utf8]{inputenc}
\usepackage[T1]{fontenc}

\DeclareUnicodeCharacter{2212}{-}
\DeclarePairedDelimiter\set\{\}

\usepackage{xr} 

\begin{document}

\include{MyCommand}
\newcommand{\BR}[1]{{\color{red}{#1}}}
\newcommand{\TS}[1]{{\color{blue}{#1}}}
\newcommand{\HB}[1]{{\textcolor[rgb]{0.4,0.4,0.8} {#1}}}
\newcommand{\AD}[1]{{\textcolor[HTML]{FF8200}{#1}}}

\newcommand{\avg}[1]{\expval{#1}}
\newcommand{\ren}{R{\'e}nyi~}

\preprint{AIP/123-QED}

\title{Disentangling the Physics of the Attractive Hubbard Model via the Accessible and Symmetry-Resolved Entanglement Entropies}

\author{Tong Shen}
 \affiliation{Department of Chemistry, Brown University, Providence, RI 02912
 }
  \author{Hatem Barghathi}
 \affiliation{Department of Physics and Astronomy, University of Tennessee, Knoxville, TN 37916
 } 
 \author{Adrian Del Maestro}
 \affiliation{Department of Physics and Astronomy, University of Tennessee, Knoxville, TN 37916
 }
\email{Author to whom correspondence should be addressed: Adrian.DelMaestro@utk.edu}
 \affiliation{Min H.~Kao Department of Electrical Engineering and Computer Science, University of Tennessee, Knoxville, TN 37996, USA}
 \affiliation{Institute for Advanced Materials \& Manufacturing, University of Tennessee, Knoxville, TN 37996, USA}
\author{Brenda M. Rubenstein}
\email{Author to whom correspondence should be addressed: brenda\_rubenstein@brown.edu.}
 \affiliation{Department of Chemistry, Brown University, Providence, RI 02912
 }
  \affiliation{Department of Physics, Brown University, Providence, RI 02912
 }

\date{\today}
\begin{abstract}
The complicated ways in which electrons interact in many-body systems such as molecules and materials have long been viewed through the lens of local electron correlation and associated correlation functions. However, quantum information science has demonstrated that more global diagnostics of quantum states, like the entanglement entropy,  can provide a complementary and clarifying lens on electronic behavior. One particularly useful measure that can be used to distinguish between quantum and classical sources of entanglement is the accessible entanglement, the entanglement available as a quantum resource for systems subject to conservation laws, such as fixed particle number, due to superselection rules. In this work, we introduce an algorithm and demonstrate how to compute accessible and symmetry-resolved entanglements for interacting fermion systems.  This is accomplished by combining an incremental version of the swap algorithm with a recursive Auxiliary Field Quantum Monte Carlo algorithm recently developed by the authors. We apply these tools to study  the pairing and charge density waves exhibited in the paradigmatic attractive Hubbard model via entanglement. We find that the particle and spin symmetry-resolved entanglements and their related full probability distribution functions show very clear - and unique - signatures of the underlying electronic behavior even when those features are less pronounced in more conventional correlation functions. Overall, this work provides a systematic means of characterizing the entanglement within quantum systems that can grant a deeper understanding of the complicated electronic behavior that underlies quantum phase transitions and crossovers in many-body systems.
\end{abstract}

\keywords{Entanglement Entropy, Accessible Entanglement, Symmetry-Resolved Entanglement, Auxiliary Field Quantum Monte Carlo, Attractive Hubbard Model}

\maketitle

\section{Introduction}


Traditionally, the complexity of condensed matter systems has been probed through the lens of correlation by focusing on local order parameters and observables like spin and charge correlation functions that are often directly accessible from scattering experiments \cite{Ament:2011ri,Bramwell:2014wy,lovesey1986theory}. While local correlations can characterize many important features of quantum systems, such as their magnetic ordering and charge distributions, in recent years, researchers have increasingly begun to view quantum systems through the lens of quantum information theory \cite{Osterloh:2002ox, Vidal:2003pw, Dowling:2006rz, Laflorencie:2016vd, Nielsen_Chuang}, especially in the absence of local symmetry breaking \cite{Kitaev:2006ly}.  This new perspective emphasizes information metrics such as the entanglement entropy \cite{Amico:2008en,Horodecki2009}, mutual information \cite{Valdez_PRL_2017,Walsh_PRL_2019}, and fidelity susceptibility \cite{Wang_PRX_2015,Gu_PRB_2008}, providing complementary non-local measures to supplement conventional probes.  These approaches have also found traction across fields where they have been used to distinguish between static and dynamic contributions to the correlation energy in molecules \cite{Plasser_JCP_2016,boguslawski_entanglement_2012,stein_automated_2016,huang_entanglement_2005,RISSLER2006519} and assess correlation during bond formation and breaking processes \cite{Boguslawski_IJQC_2015,boguslawski_orbital_2013}. In condensed matter, quantum information metrics have provided key insights into phases and phase transitions that do not fall within the traditional symmetry-breaking framework, including topological orders and related transitions \cite{Kitaev:2006ly,Levin:2006qz,Isakov:2012ku,Wen:2019oe, Hu:2020qx}. 

Among the various notions in quantum information theory, entanglement is amongst the most foundational, encapsulating the non-classical correlations in a quantum state \cite{Horodecki2009,LAFLORENCIE20161}. For a pure state, it can be quantified by the von Neumann and R\'enyi entanglement entropies.  For a many-body quantum system described by such a pure state that has non-zero entanglement, it is natural to ask if that entanglement could be used as a resource for quantum information processing applications.  This question can be addressed via the accessible entanglement, which quantifies the ``useful'' entanglement that can be transferred to a quantum register in the presence of conservation laws \cite{Wiseman:2003ei, Bartlett:2003ec, Barghathi:2018rg,Barghathi:2019db} and can even be measured in experiments in ultracold atoms \cite{Melko:2016tv,Lukin:2019dy}. This is an important problem, as in many physical systems, certain properties are conserved due to underlying symmetries or other physical constraints. Such conservation laws restrict the set of possible local operations that can be performed on a system, imposing so-called superselection rules \cite{Bartlett:2003ec,Kitaev:2004ec,Schuch:2004vj,Bartlett:2007gb} that limit the amount of entanglement that can be extracted via local operations and classical communication.  For example, for particle number conservation, the corresponding superselection rule precludes the extraction of quantum information generated from particle number fluctuations alone, thereby limiting the accessible entanglement to less than the von Neumann or R\'enyi entanglement entropies \cite{Klich:2008se,Barghathi:2018rg}. In addition to its importance as an experimentally relevant bound on the extractable entanglement, accessible entanglement, when compared with the full entanglement, is more sensitive to the underlying particle statistics even in the absence of interactions \cite{Barghathi:2018rg} and may display signals across quantum phase transitions \cite{barghathi2020theory,Barghathi:2019db} in interacting systems. However, the definition of accessible entanglement hinges on the symmetry resolution of the entanglement that is enforced by the presence of the corresponding conservation laws \cite{Goldstein:2018,Benatti:2020,Turkeshi:2020,Faiez:2020,Horvath:2020,Bonsignori:2020,Horvath:2021,Murciano:2021,Estienne:2121,Foligno:2023,Northe:2023,Fossati:2023,Monkman:2023,deGroot:2020,Dana:2020,Cornfeld:2019}. Thus, the entanglement symmetry resolution that defines accessible entanglement can be employed to gain more insight into the entanglement characteristics of a target state. 

While these different approaches to quantifying entanglement may readily be quantified for non-interacting systems \cite{Tan:2020FF,Barghathi:2018rg,Murciano:2020,Monkman:2023}, in one dimension via tensor network approaches \cite{Barghathi:2019db}, or via exact diagonalization for small interacting systems \cite{Barghathi:2019db,Melko:2016tv}, it is only in the past decade that various quantum Monte Carlo techniques have been devised to compute these quantities \cite{Hastings_PRL_2010,Humeniuk_PRB_2012,McMinis_PRB_2013,grover2013entanglement, Tubman_PRB_2014,Herdman:2014cp,Assaad_PRB_2015,Drut_PRB_2015,CasianoDiaz:2023pi} that mostly rely on the swap algorithm \cite{Calabrese:2004ll}. The swap algorithm forgoes the need for full state tomography and the exact computation of density matrices, recasting entanglement as the expectation value of a local operator that can be measured as the ratio of special partition functions.  However, this usually comes at the cost of simulating additional copies of the system (so-called replicas) and thus exacerbates the inherent difficulties in accurately modeling large interacting fermionic many-body systems beyond one spatial dimension.  Algorithms for computing the accessible (or symmetry-resolved) entanglement of interacting systems have proven even more challenging, as they necessitate the computation of properties subject to conservation laws which lead to global constraints.  For the case of particle number conservation, this requires performing simulations in the canonical ensemble \cite{Shen_JCP_2020,Shen_PRE_2023,CasianoDiaz:2023pi} and most previous numerical work has focused on non-interacting \cite{Klich:2008se,Barghathi:2018rg,Parez_2021,Parez_PRB_2021,Kiefer_Emmanouilidis:2020} or low-dimensional interacting systems \cite{Barghathi:2019db,Laflorencie:2014}. However, these previous studies have broken considerable new ground, shedding light on how fluctuations and full counting statistics contribute to the entanglement of strongly interacting quantum matter.  However, a more complete picture of accessible and symmetry-resolved entanglement near phase transitions and crossovers in interacting systems beyond one dimension and in the presence of multiple competing conservation laws is still lacking. 

In this manuscript, we introduce a formalism and algorithms for quantifying the accessible and symmetry-resolved entanglements of systems of interacting spinful fermions. We show in particular how to compute spin- and particle-resolved entanglements and their related particle number probability distribution functions using Auxiliary Field Quantum Monte Carlo (AFQMC) \cite{White_PRB_1989,Zhang_PRB_1997} that can scale to systems sizes of many tens to hundreds of sites.  A strength of this approach is the ability to implement designer partitions of the many-body Hilbert space not constrained by the locality needed in tensor network approaches.  This allows us to disentangle the relative contributions of configurations, local and non-local constraints, and fluctuations of the interacting charge and spin degrees of freedom to entanglement. To do so, we employ an incremental version of the swap algorithm \cite{Hastings_PRL_2010} and combine it with a recursive algorithm recently developed by the authors \cite{Shen_PRE_2023} that grants access to subsystem entanglements corresponding to specific conserved quantum numbers. We leverage these measures to study emergent electronic behavior in the attractive Hubbard model, observing signatures of pairing and charge density waves in not only the computed entanglement entropies, but also in the particle number and magnetization probability distributions. Importantly, we show that these entanglement signatures are often more pronounced than those appearing in more conventionally studied pair correlation functions, and that they can exhibit multiple signs of ordering at once. The computed symmetry-resolved entanglements highlight the specific charge and spin sectors that contribute most to the entanglement and system fluctuations. While the attractive Hubbard model is paradigmatic and well understood, we use it as a benchmark case to demonstrate that entanglement measures can provide additional information into ordering, pairing, and quantum phenomena in many-body systems with nontrivial electronic behavior. 

We begin with a discussion of the theoretical formalism that underlies the calculation of the R{\'e}nyi entanglement entropy and its accessible and symmetry-resolved counterparts in Section \ref{formal}. We then proceed to describe how these measures can be computed using a recursive AFQMC method that employs replicas of ensembles to calculate entanglements and other quantum information metrics. After presenting these formal and numerical details, in Section \ref{res}, we illustrate the effectiveness of these measures by analyzing the pairing and charge density waves in the attractive Hubbard model on quarter and checkerboard partitions of the lattice at two different filling fractions. We conclude with a discussion of potential improvements to and advancements upon the algorithms presented here in Section \ref{conc}. 

\section{Formalism}
\label{formal}

The efficient AFQMC algorithms presented herein to study spin- and charge-resolved entanglement entropies are based on the computation of \ren entanglement entropies.  We thus begin by introducing the formalism underlying the \ren entanglement entropy, before deriving its accessible and particle- and spin-resolved variants. We then present the new and statistically-improved techniques required to converge these quantities within a quantum Monte Carlo framework. 

\subsection{\ren Entanglement Entropy}

We focus on a $d$-dimensional finite lattice of size $L^d = N_s$, where $N_s$ denotes the number of sites that can be occupied by some fixed number, $N$, of interacting spinful fermions with filling fraction $N/N_s$. Given a pure state $\rho=\vert\Psi\rangle\langle\Psi\vert$ of a quantum many-body system, the entanglement that exists between a partition $A$ and its complement $\bar{A}$ can be quantified by calculating the reduced density matrix
\begin{equation}
    \rho_{A}=\Tr_{\bar{A}}\rho,
\end{equation}
via a partial trace over the degrees of freedom in the complementary partition $\bar{A}$. While there is no single unique measure of non-classical correlations, the entanglement can be quantified through the $\alpha$th \ren entanglement entropy 
\begin{equation}
S_{\alpha}(\rho_A)=\frac{1}{1-\alpha}\ln\Tr\rho_A^\alpha,
\end{equation}
where $\alpha$ is the \ren index \cite{Nielsen_Chuang,Barghathi:2018rg}.  In this paper, we focus on the $\alpha=1$ (von Neumann) and $\alpha=2$ \ren entanglement entropies because they can be most readily computed and characterized.

\subsection{Symmetry-Resolved Entanglement Entropy}
Despite its usefulness, $S_{\alpha}(\rho_A)$ summarizes the entanglement information encoded in $\rho_A$ in just a single number that may not convey the full complexity of the entanglement of a given system. In the presence of a conserved quantity (conserved due to superselection rules which restrict certain quantum superpositions), measured by the operator $\hat{Q}$ such that $\left[\rho,\hat{Q}\right]=0$, one can gain more insight into the entanglement structure by determining the entanglement in each of the conserved sectors. In such circumstances, the reduced density matrix $\rho_A$ has a block-diagonal structure, where each block can be directly associated with the quantity $q$ contained in partition $A$.  Common physical examples include the conservation of the total number of particles or magnetization which cannot change through local fluctuations. Accordingly the reduced density matrix for partition $A$ can be decomposed as
\begin{equation}
\rho_A=\sum_q \hat{\Pi}_{A_q}\rho_A\hat{\Pi}_{A_q},
\end{equation}
where $\hat{\Pi}_q$ is a projection operator that fixes $q$ in $A$. Consequently, $\rho_A^\alpha$ is also block-diagonal in $q$, permitting the resolution of $\Tr\rho_A^\alpha$ over $q$ as
\begin{equation}
e^{\left(1-\alpha\right)S_\alpha(\rho_A)}=\Tr\rho_A^\alpha=\sum_q \Tr\hat{\Pi}_{A_q}\rho_A^\alpha\hat{\Pi}_{A_q}.
\label{def:Trace-resolve}
\end{equation}
Thus, the relative contribution of sector $q$ to $\Tr\rho_A^\alpha$ can be captured by the effective probability distribution
\begin{equation}
P_{q,\alpha}=\frac{\Tr\hat{\Pi}_{A_q}\rho_A^\alpha\hat{\Pi}_{A_q}}{\Tr\rho_A^\alpha}, \label{def:P_qalpha}
\end{equation}
which converges to the distribution $P_{q}=\Tr\hat{\Pi}_{A_q}\rho_A\hat{\Pi}_{A_q}$ in the limit $\alpha\to1$.

While $P_{q,\alpha}$ provides information about the contributions of the different $q$ sectors to $\Tr\rho_A^\alpha$, it does not directly describe the entanglement content in each of the $q$ sectors. Fortunately, this entanglement can be obtained by normalizing each projected density matrix as $\rho_{A_q}=[\hat{\Pi}_{A_q}\rho_A\hat{\Pi}_{A_q}]P_q^{-1}$
and applying the \ren measure to each of them, where the symmetry-resolved entanglement is defined by
\begin{equation}
S_{\alpha}(\rho_{A_q})=\frac{1}{1-\alpha}\ln\Tr\rho_{A_q}^\alpha. \label{S_alpha}
\end{equation}
Employing the fact that the definitions of both $P_{q,\alpha}$ and $S_{\alpha}(\rho_{A_q})$ depend on $\Tr\hat{\Pi}_{A_q}\rho_A^\alpha\hat{\Pi}_{A_q}$, we can write 
\begin{equation}
    P_{q,\alpha}=P_q^{\alpha}e^{\left(\alpha-1\right)\qty[S_\alpha(\rho_A)-S_{\alpha}(\rho_{A_q})]}.
\label{P_alpha_Delta}
\end{equation}
This shows that $P_{q,\alpha}$ depends on both $P_q$ and the differences between the total entanglement entropy and the symmetry-resolved entanglement entropy where we define the difference
\begin{equation}
\Delta S_{\alpha}(\rho_{A_q}):=S_{\alpha}(\rho_{A}) -S_{\alpha}(\rho_{A_q}),
\end{equation}
as an entanglement measure resolving the $q$ sector. It can be calculated as
\begin{equation}
    \Delta S_{\alpha}(\rho_{A_q}) = \frac{1}{\alpha-1} \ln \frac{P_{q,\alpha}}{P_{q,1}^{\alpha}}
    \label{eq:entang_diff}
\end{equation}
and the symmetry-resolved entanglement entropy of different conserved quantities can be obtained in this way. Below we focus on two of the most commonly conserved quantities: fixed total particle number and fixed total spin (magnetization). 

\subsection{Accessible Entanglement Entropy}
Wiseman and Vaccaro introduced the idea of the accessible entanglement, which refers to the quantum entanglement that can be extracted from a many-body state and then transferred to a quantum register while a superselection rule (SSR) is in place \cite{Wiseman:2003ei}. For fixed total particle number, the accessible entanglement is defined as the weighted sum of the symmetry-resolved entanglement entropies introduced in the previous section where the quantity, $q$, is taken to be the local particle number in partition $A$, $n_A$.  For the von Neumann entropy, the accessible entanglement has a simple definition
\begin{equation}
    S_1^{\rm acc}(\rho_A) = \sum_q P_q S_1(\rho_{A_q}),
\end{equation}
where $P_q$ is the probability of partition $A$ having $n_A = q$ particles,
\begin{equation}
    P_q  = P_{q,1} = \Tr(\hat{\Pi}_{A_q} \rho_A \hat{\Pi}_{A_q}).
\end{equation}
For $\alpha>1$, the accessible \ren entropy has been shown to take the form \cite{Barghathi:2018rg}
\begin{equation}
    S_{\alpha}^{\rm acc}(\rho_A) = \frac{\alpha}{1-\alpha} \ln\qty[\sum_q P_q e^{\frac{1-\alpha}{\alpha}S_{\alpha}(\rho_{A_q})}], \label{eqn:S2Acc_Renyi}
\end{equation}
which reduces to $S^{\rm acc}_1$ in the limiting case $\alpha\rightarrow 1$. Although Eq.~\eqref{eqn:S2Acc_Renyi} has a rather complicated form, it can be simplified and associated with $S_\alpha(\rho_A)$ via a decomposition \cite{Klich:2008se,Barghathi:2018rg}
\begin{equation}
    S_\alpha(\rho_A) = S_{\alpha}^{\rm acc}(\rho_A) + H_{1/\alpha}(\set{P_{q,\alpha}}), \label{eqn:AccRenyiDecomposition}
\end{equation}
where $H_{1/\alpha}$ denotes a generalized Shannon entropy of the probability distribution $P_{q,\alpha}$ with index $1/\alpha$ that captures the contribution to the entropy from fluctuations \cite{Barghathi:2018rg}
\begin{equation}
    H_{1/\alpha}(\set{P_{q, \alpha}}) = \frac{1}{1-\alpha} \ln \sum_q (P_{q, \alpha})^{1/\alpha}. \label{eqn:Shannon}
\end{equation}
While $H_{1/\alpha}$ can be defined as a function of any probability distribution, in the remainder of the manuscript, we focus on $H_{1/\alpha}(\{P_{q,\alpha}\})$ whenever referring to or plotting $H_{1/\alpha}$. Note that Eq.~\eqref{eqn:AccRenyiDecomposition} demonstrates that the overall entanglement entropy may be decomposed into the sum of an accessible component and the generalized Shannon entropy, which can be viewed as the entropy stemming from ``classical'' $q$-fluctuations. This implies that the accessible entanglement may be accessed by subtracting the Shannon entropy from the entanglement entropy, as we do in the following.

While the accessible and symmetry-resolved entanglements may be defined and computed for any value of $\alpha$, here, we focus on the $\alpha=2$ case and the related R{\'e}nyi-2 entropies, which are easier to obtain numerically.

\subsection{Computing Symmetry-Resolved and Other Entanglement Entropies Using Auxiliary Field Quantum Monte Carlo}

In this section, we illustrate how to compute $S_2(\rho_A)$ and $P_{q, \alpha}$ for $\alpha=1,2$ (and their symmetry-resolved counterparts) using the AFQMC method, thereby gaining access to the symmetry-resolved and accessible entanglement measures for strongly interacting fermionic systems.  Since we do not employ any constraints, the AFQMC we use is equivalent to Determinant Quantum Monte Carlo \cite{Loh_PRB_1990}. 

\subsubsection{Computing the R{\'e}nyi-2 Entropy}
As demonstrated by Grover \cite{grover2013entanglement}, \ren entanglement entropies of interacting fermion systems can be computed using QMC methods. One such method that has been shown to be highly accurate for strongly-correlated lattice models and is particularly well-suited for computing the associated entropy is AFQMC \cite{Hirsch_PRB,Zhang_PRL_2003,Zhang_PRB_1997}. In AFQMC, a Hubbard-Stratonovich (HS) Transformation \cite{stratonovich,hubbard,Hirsch_PRB_1983} is introduced to decouple the many-body interactions, such that the imaginary-time propagator acts on a trial wavefunction $\vert \phi_T\rangle$  over an imaginary projection time $\Theta$ and can be written as a path integral over auxiliary fields ${\bf s}$
    \begin{equation}
        e^{-\Theta \hat{H}} \vert \phi_T\rangle = \int \mathcal{D}{\bf s}\,p_{\bf s} \hat{U}_{\bf s}  \vert \phi_T\rangle.
        \label{hs}
    \end{equation}
The effective one-body propagator $\hat{U}_{\bf s}$ is determined from the underlying Hamiltonian \cite{Motta_WIRES,Zhang_PRB_1997} and $p_{\bf s}$ is a probability measure. In our modeling below, we use a discrete HS transform for the Hubbard Hamiltonian, where a local two-body operator propagated by a imaginary time step $\Delta\tau$ can be decomposed as
\begin{equation}
    e^{-\Delta\tau U \hat{n}_{\uparrow} \hat{n}_{\downarrow}} = \frac{1}{2} \sum_{s=\pm 1} e^{\gamma s (\hat{n}_{\uparrow} - \hat{n}_{\downarrow})}e^{-\frac{\Delta\tau U}{2} (\hat{n}_{\uparrow} + \hat{n}_{\downarrow})},
\end{equation}
with $\cosh(\gamma)=e^{\frac{\Delta\tau U}{2}}$. Here, $p_{\bf s}$ becomes a Bernoulli distribution, ${\rm B}(\frac{1}{2})$, and $\hat{U}_{\bf s}$ is a combined propagator from the above decomposition and the remaining one-body kinetic part of the Hamiltonian.  The exact ground state wavefunction is obtained in the limit of infinite projection time $\Theta$ as $\vert \phi_0\rangle = \lim_{\Theta\rightarrow\infty} e^{-\Theta \hat{H}} \vert \phi_T\rangle$. It should be noted that Eq.~\eqref{hs} implies that any two-body propagator can be re-expressed as an integral over weighted one-body propagators, which signifies that properties of an interacting system, like the entanglement, can be obtained by integrating over their non-interacting counterparts appropriately, a fact that we will make extensive use of below. 

To facilitate derivations in subsequent sections, we recast the ground state average of any operator $\hat{O}$ in a form reminiscent of a thermal average
    \begin{equation}
        \langle \hat{O} \rangle = \frac{\int \mathcal{D}{\bf s}\, Z_{\bf s} \expval{O}_{\bf s}}{Z}. \label{eqn:obs_general}
    \end{equation}
    In contrast with a typical thermal average, where statistical weights are given by the partition function, $Z_{\bf s}$ and $Z$ represent the overlaps of the trial wavefunction with itself after propagation by $\hat{U}_{\bf s}$ and $e^{-\Theta \hat{H}}$, respectively, i.e.,
    \begin{equation}
        Z_{\bf s} := \frac{\langle \phi_T\vert \hat{U}_{\bf s} \vert \phi_T\rangle}{\langle \phi_T\vert \phi_T\rangle}, Z := \frac{\langle \phi_T\vert e^{-\Theta \hat{H}} \vert \phi_T\rangle}{\langle \phi_T\vert \phi_T\rangle}.
    \end{equation}
    Note that $Z_{\bf s}/Z$ can be interpreted as the probability distribution for the field configuration $\bf s$, but could take negative values, which gives rise to the sign problem \cite{Loh_PRB_1990}.

    Grover proved \cite{grover2013entanglement} that the reduced density matrix (RDM) can be decomposed into the same form as Eq.~\eqref{eqn:obs_general}
    \begin{equation}
        \rho_A = \frac{\int \mathcal{D}{\bf s}\, Z_{\bf s} \rho_{A, {\bf s}}}{Z},
    \end{equation}
where, for each configuration $\bf s$, one obtains an RDM associated with an entanglement Hamiltonian $\hat{H}_{A}$ that only contains one-body, local terms
    \begin{equation}
        \rho_{A, \textbf{s}} = \det(\mathbb{I} - G^{\bf s}_{A}) e^{-\hat{H}_{A}^{\bf s}} \label{eqn:FreeFermionRDM},
    \end{equation}
    where $\mathbb{I}$ is the identity operator and 
\begin{equation}
    \hat{H}_{A}^{\bf s} = \hat{c}^{\dag} \log [(G^{\bf s}_{A})^{-1} - \mathbb{I}] \hat{c}. \label{eqn:1-RDM_HSForm}
\end{equation}
Here, $\hat{c}_j^{\dagger}(\hat{c}_j^{\phantom \dagger})$ are fermionic creation(destruction) operators such that $\qty{\hat{c}_i^{\phantom \dagger},\hat{c}_j^{\dagger}}=\delta_{ij}$ and  the equal-time Green's function $(G^{\bf s}_{A})_{ij} = \expval{\hat{c}^{\dag}_j \hat{c}_i^{\phantom \dagger}}_{\bf s}$, is defined such that $i,j$ are restricted to sites in subsystem $A$. 

As a result, the R{\'e}nyi-2 entropy $S_2(\rho_A)=-\ln\Tr\,[(\rho_A)^2]$ can be evaluated using two independent replicas as
\begin{equation}
    e^{-S_2(\rho_A)} = \frac{\int \mathcal{D}{{\bf s}_1} \mathcal{D}{\vb{s}_2}\, Z_{{\bf s}_1} Z_{{\bf s}_2} \det g_A^{{{\bf s}_1}, {{\bf s}_2}}}{Z^2}, \label{eqn:S2_MC} 
\end{equation}
where the Grover matrix $g_A^{{{\bf s}_1}, {{\bf s}_2}}$ is defined as a functional of $G^{{\bf s}_1}_{A}$ and $G^{{\bf s}_2}_{A}$, $g_A^{{{\bf s}_1}, {{\bf s}_2}}:= G^{{\bf s}_1}_{A} G^{{\bf s}_2}_{A} + (\mathbb{I} - G^{{\bf s}_1}_{A})(\mathbb{I} - G^{{\bf s}_2}_{A})$. Previous studies \cite{broecker2014renyi, pan2023stable} have shown that this independent replica structure can result in significant statistical fluctuations, as infrequent pairs of $({\bf s}_1, {\bf s}_2)$ often lead to large values for $\det g_A^{{{\bf s}_1}, {{\bf s}_2}}$ and contribute significantly to the evaluation of Eq.~\eqref{eqn:S2_MC}. To suppress these large statistical fluctuations, various algorithms have been introduced that show enhancements over Grover's initial method, including the swap \cite{hastings2010measuring,broecker2014renyi,broecker2016numerical} and incremental \cite{d2020entanglement,d2022universal} algorithms. 

In this work, we utilize a recently proposed incremental algorithm  \cite{da2023controllable} that is both more numerically stable than Grover's method and more efficient than other incremental algorithms, and extend it to allow for the computation of symmetry-resolved quantum information measures.  This approach is analogous to thermodynamic integration, where an auxiliary parameter $0\leq \lambda \leq 1$ is introduced. One then defines a function $\mathcal{Z}(\lambda)$ via
\begin{eqnarray}
    \mathcal{Z}(\lambda) &=& \int \mathcal{D}{{\bf s}_1} \mathcal{D}{{\bf s}_2}\, \mathcal{Z}({{\bf s}_1}, {{\bf s}_2}, \lambda) \nonumber\\
    &=& \int \mathcal{D}{{\bf s}_1} \mathcal{D}{{\bf s}_2} Z_{{\bf s}_1} Z_{{\bf s}_2} (\det g_A^{{{\bf s}_1}, {{\bf s}_2}})^{\lambda}, \label{eqn:incremental_def}
\end{eqnarray}
such that the R{\'e}nyi-2 entropy can be computed incrementally across a grid that ranges between $\lambda=0$ and $\lambda=1$:
\begin{eqnarray}
    e^{-S_2(\rho_A)} &=& \frac{\mathcal{Z}(\lambda=1)}{\mathcal{Z}(\lambda=0)} \nonumber\\
    &=& \frac{\mathcal{Z}(\Delta\lambda)}{\mathcal{Z}(0)} \frac{\mathcal{Z}(2\Delta\lambda)}{\mathcal{Z}(\Delta\lambda)} \cdots \frac{\mathcal{Z}(1)}{\mathcal{Z}(1-\Delta\lambda)}. \label{eqn:incremental_prods}
\end{eqnarray}
Defining $N_\lambda = 1 / \Delta\lambda$, each auxiliary ratio can be computed with QMC as
\begin{eqnarray}
    \frac{\mathcal{Z}\qty((k+1)\Delta\lambda)}{\mathcal{Z}(k\Delta\lambda)}\!\! &=& \!\!\frac{\int \mathcal{D}{\vb{s}_1} \mathcal{D}{\vb{s}_2} \mathcal{Z}({\vb{s}_1}, {\vb{s}_2}, k\Delta\lambda) (\det g_A^{{\vb{s}_1}, {\vb{s}_2}})^{1/N_\lambda}}{\mathcal{Z}(k\Delta\lambda)} \nonumber\\
                                                                                &=& \expval{(\det g_A^{{\vb{s}_1}, {\vb{s}_2}})^{1/N_\lambda}}_{\mathcal{Z}(k\Delta\lambda)} \label{eqn:incremental_estimator},
\end{eqnarray}
where sample fluctuations are suppressed by introducing a new estimator $(\det g_A^{{{\bf s}_1}, {{\bf s}_2}})^{1/N_{\lambda}}$. A suitable choice of $N_{\lambda}$, which can be determined through several trial runs, makes the fluctuations of this estimator of the order 1 such that $\delta[(\det g_A^{{{\bf s}_1}, {{\bf s}_2}})^{1/N_{\lambda}}] \sim 1$ \cite{da2023controllable}. In Appendices \hyperref[replica]{A} and \hyperref[swap_and_incremental]{B}, we analyze the accuracy and sampling efficiency of this incremental algorithm and demonstrate that it exhibits reduced statistical noise compared to the more conventional swap algorithm.

\subsubsection{Probability Distributions for Symmetry Sectors}

Since measuring $P_{q, \alpha}$ necessitates conducting QMC simulations with $\alpha$ replicas, evaluating $P_{q, 1}$ is relatively straightforward as it can be performed within the regular AFQMC framework for a single replica
\begin{equation}
    P_{q,1} = \frac{\int \mathcal{D}{\bf s}\, Z_{\bf s} \expval{P_{q,1}}_{\bf s}}{Z}.
\end{equation}
Importantly, computing the estimator $\expval{P_{q,1}}_{\bf s}$ in the presence of a conservation law for different $q$ values necessitates a canonical ensemble algorithm that can determine observables for fixed $q$ values. To accomplish this task, we leverage the recursive canonical ensemble algorithm we have previously developed \cite{Shen_PRE_2023}, which enables the measurement of this estimator for all allowable $q$ values simultaneously using the recursive relation
\begin{equation}
    \expval{P_{q,1}}_{\bf s} = p^{(1)}_i \expval{P^{\set{\lambda^{(1)}}\backslash \lambda^{(1)}_i}_{q-1,1}}_{\bf s} + (1 - p^{(1)}_i) \expval{P^{\set{\lambda^{(1)}}\backslash \lambda^{(1)}_i}_{q,1}}_{\bf s}, \label{eqn:P_q1_recursion}
\end{equation}
where $\set{\lambda^{(1)}}$ is the exponential of the effective entanglement spectrum under the fields $\bf s$, which can be obtained by diagonalizing the propagator matrix $e^{- \mathbf{H}^{\bf s}_{A}}$
\begin{equation}
{\rm Diag}(\set{\lambda^{(1)}}) = {\bf Q}_1^{-1} \frac{G^{\bf s}_A}{\mathbb{I} - G^{\bf s}_A} {\bf Q}_1^{\phantom 1}.
\end{equation}
Here, $p^{(1)}_i$ is the corresponding level occupancy for the $i$th eigenvalue, $p^{(1)}_i = \frac{\lambda^{(1)}_i}{1+\lambda^{(1)}_i}$.

Similar to the calculation of $S_2(\rho_A)$, resolving $P_{q,2}$ requires two replicas. We first rewrite Eq.~\eqref{def:P_qalpha} for $\alpha=2$ as the ratio of two partition functions
\begin{equation}
    P_{q, 2} := \frac{Z^{(2)}_{A_q}}{Z^{(2)}_A}, \label{eqn:Pq2_ratioform}
\end{equation}
where $Z^{(2)}_{A_q}$ is defined as the normalization of the squared reduced density matrix projected onto the $q$-particle symmetry sector, such that
\begin{equation}
    \Tr[\Pi_{A_q} \rho^{2}_A \Pi_{A_q}] = \frac{Z^{(2)}_{A_q}}{Z^2}.
\end{equation}
Under the assumption that there exists an auxiliary field decomposition for this normalization, $Z^{(2)}_{A_q} = \int \mathcal{D}{{\bf s}_1} \mathcal{D}{{\bf s}_2}\,Z^{(2)}_{A_q}({{\bf s}_1}, {{\bf s}_2})$, then Eq.~\eqref{eqn:Pq2_ratioform} can also be expressed in terms of auxiliary fields
\begin{eqnarray}
    P_{q, 2} &=& \frac{1}{Z^{(2)}_A} \int \mathcal{D}{{\bf s}_1} \mathcal{D}{{\bf s}_2}\, Z^{(2)}_{A_q}({{\bf s}_1}, {{\bf s}_2}) \nonumber\\
    &=& \frac{1}{Z^{(2)}_A} \int \mathcal{D}{{\bf s}_1} \mathcal{D}{{\bf s}_2}\, \frac{Z^{(2)}_{A_q}({{\bf s}_1}, {{\bf s}_2})}{Z^{(2)}_{A}({{\bf s}_1}, {{\bf s}_2})}Z^{(2)}_{A}({{\bf s}_1}, {{\bf s}_2}),
\end{eqnarray}
which enables QMC sampling via the estimator
\begin{equation}
    \langle P_{q, 2}({{\bf s}_1}, {{\bf s}_2}) \rangle_{Z^{(2)}_A} := \Biggl \langle \frac{Z^{(2)}_{A_q}({{\bf s}_1}, {{\bf s}_2})}{Z^{(2)}_A({{\bf s}_1}, {{\bf s}_2})} \Biggr \rangle_{Z^{(2)}_A}. \label{eqn:alpha_prob_estimator}
\end{equation}
To evaluate $P_{q, 2}({{\bf s}_1}, {{\bf s}_2})$, we need to project the product of the field-dependent RDMs 
\begin{equation}
    \rho_{A, {{\bf s}_1}} \rho_{A, {{\bf s}_2}} = \det(\mathbb{I} - G^{{\bf s}_1}_{A}) \det(\mathbb{I} - G^{{\bf s}_2}_{A}) e^{-(\hat{H}_{A}^{{\bf s}_1} + \hat{H}_{A}^{{\bf s}_2})}
\end{equation}
onto the symmetry sector $q$, according to Eq.~\eqref{def:P_qalpha}. Similar to Eq.~\eqref{eqn:P_q1_recursion}, this can be accomplished recursively via
\begin{align}
    \expval{P_{q,2}({{\bf s}_1}, {{\bf s}_2})}_{Z^{(2)}_A} & =  p^{(2)}_i \expval{P^{\set{\lambda^{(2)}}\backslash \lambda^{(2)}_i}_{q-1,2} ({{\bf s}_1}, {{\bf s}_2})}_{Z^{(2)}_A} \nonumber\\
    & + (1 - p^{(2)}_i) \expval{P^{\set{\lambda^{(2)}}\backslash \lambda^{(2)}_i}_{q,2} ({{\bf s}_1}, {{\bf s}_2})}_{Z^{(2)}_A}, \nonumber \\
    \label{eqn:P_q2_recursion}
\end{align}
based on the eigendecomposition of the matrix $e^{- (\mathbf{H}^{{\bf s}_1}_{A} + \mathbf{H}^{{\bf s}_2}_{A})}$
\begin{equation}
    {\rm Diag}(\set{\lambda^{(2)}}) = {\bf Q}_2^{-1} \frac{G^{\vb{s}_1}_A}{\mathbb{I} - G^{\vb{s}_1}_A} \frac{G^{\vb{s}_2}_A}{\mathbb{I} - G^{\vb{s}_2}_A} {\bf Q}_2^{\phantom 1},
\end{equation}
and the corresponding level occupancy $p^{(2)}_i = \frac{\lambda^{(2)}_i}{1+\lambda^{(2)}_i}$.

\section{Illustrative System: The Attractive Hubbard Model}

To illustrate how symmetry-resolved entanglements can provide additional
information beyond the unresolved \ren entanglement, we examine how they behave
in different parameter regimes of the attractive (negative-$U$) Hubbard model
\cite{Scalettar:1989th,Fontenele:2022ex}. This model describes the pairing of
electrons of different spins and may capture some of the physics 
occurring in superconductors and Bose-Einstein condensates (BEC). The Hamiltonian
of the attractive Hubbard Model (AHM) may be expressed as
\begin{equation}
\hat{H} = -t\!\!\! \sum_{\langle i,j\rangle,\sigma}\!\! \qty( \hat{c}_{i,\sigma}^{\dagger} \hat{c}_{j,\sigma} + {\rm H.c.}) -|U| \sum_{i} (\hat{n}_{i,\uparrow}-\frac{1}{2}) (\hat{n}_{i,\downarrow} - \frac{1}{2}),
\label{eq:AHM}
\end{equation}
where $\hat{c}_{i,\sigma}^{\dagger}(\hat{c}_{j,\sigma}^{\phantom\dagger})$ are
anti-commuting fermionic creation(annihilation) operators such that
$\hat{c}_i^{\phantom \dagger}\hat{c}_j^\dagger +
\hat{c}_i^{\dagger}\hat{c}_j^{\phantom\dagger} = \delta_{ij}$ and
$\hat{n}_{i,\sigma} = \hat{c}_{i,\sigma}^{\dagger} \hat{c}_{i,\sigma}^{\phantom
\dagger}$ is the local spin-resolved density. $\expval{i,j}$ denotes nearest-neighbor sites, $t$ is the hopping parameter, and $U$ represents the interaction
strength. For our ground state simulations in the canonical ensemble, it is not essential to consider the fully particle-hole symmetric form of Eq.~\eqref{eq:AHM} because there is no sign problem for the spin-balanced systems we model in this manuscript. 

Unlike the repulsive Hubbard model, the electron-electron interaction in the attractive Hubbard model is constrained to be negative ($U<0$), meaning that electrons on the same site favor pairing. We focus on two spatial dimensions on the square lattice for which there is no finite temperature phase transition in the model (\emph{i.e.},\@ $T_c = 0$) at half filling ($\expval{n_{i,\sigma}} = 1$) \cite{Scalettar:1989th}, however long-range charge density wave and pairing orders coexist in the ground state. The lack of any finite temperature ordering at half-filling can be understood in terms of the well-known mapping of Eq.~\eqref{eq:AHM} to a 2D Heisenberg model. However, away from half-filling, the existence of a finite hole density manifests as an effective external magnetic field in the spin model which effectively reduces the spin dimensionality allowing for a Kosterlitz-Thouless transition at finite $T$. Recent determinant QMC simulations \cite{Fontenele:2022ex} mapped out an interaction vs. filling phase diagram, and demonstrated the importance of carefully understanding the subtle interplay of finite size effects with the emergence of pairing and charge density wave correlations.  Motivated by this work, we take the AHM as a paradigmatic model for which entanglement measures may provide additional information over conventionally-measured correlation functions. 

In the remainder of this manuscript, we set $t=1$ and fix the total number of fermions to $N$ with zero total magnetization $M=0$ (equal contributions from both spin
flavors). We vary $|U|/t$ at $T=0$ to study how the AHM's symmetry-resolved entanglement entropies and fluctuation probability distributions vary with pairing strength and filling fraction.

\section{Results and Discussion}
\label{res}

Although the AFQMC formalism presented above is general and can operate at finite temperature, here we focus on the ground state as the entanglement measures we consider are well defined for pure states. This is achieved using an imaginary projection time of $\Theta = 18$ for the probability distributions and $\Theta = 50$ for the entanglements.  The latter is required due to the higher sensitivity of the R{\'e}nyi-2 entropy to discrepancies between the propagated trial wavefunction and the true ground state wavefunction \cite{Shen_PRE_2023}.  We have confirmed that both projection times are adequate for converging the corresponding observables to the ground state. For the trial wavefunction, we use the ground state of the Hartree-Fock Hamiltonian, which we found is more numerically stable than a BCS trial wavefunction in replica sampling and employ a Trotter step of $\Delta\tau = 0.1$, unless otherwise specified. We present results for a finite AHM with $N_s = 8 \times 8 = 64$ sites.

\subsection{Joint and Marginalized Probability Distributions for a Quarter-Partition \label{jointmargin}}

\begin{figure}
    \centering
    \includegraphics[width=0.5\textwidth]{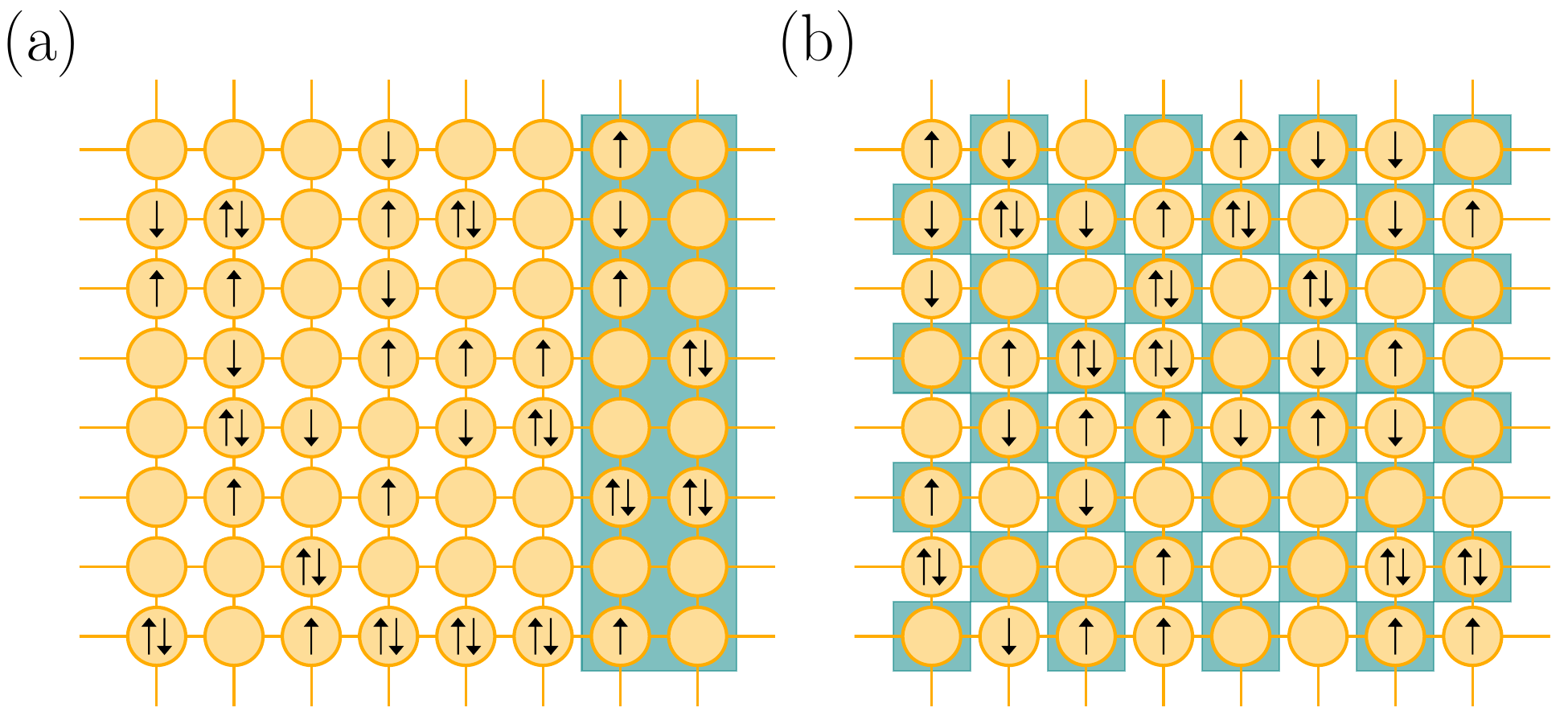}
    \caption{The two subsystem partitions (shaded regions) used in this work: (a) a contiguous $8\times2$ quarter-partition and (b) a checkerboard half-partition.}
    \label{fig:heatmap_half_filling}
\end{figure}
%
To illustrate the power of our symmetry-resolved formalism to resolve correlations among interacting fermions, we begin by analyzing the joint probability distribution functions, $P_{(N_{A\uparrow}, N_{A\downarrow}),\alpha}$, of the attractive Hubbard model at two different fillings ($\expval{n_{i,\sigma}}=1,1/2$) for a contiguous partition of the lattice containing $N_s/4$ sites (as seen in Fig.~\ref{fig:heatmap_half_filling}(a)).  The distributions $ P_{(N_{A\uparrow}, N_{A\downarrow}),\alpha}$ contain information about charge and spin fluctuations and their contribution to the total and accessible entanglement as introduced in Sec.~\ref{formal}.  In particular, $P_{(N_{A\uparrow}, N_{A\downarrow}),1}$ describes the likelihood of observing $N_{A\uparrow}$ spin-up and $N_{A\downarrow}$ spin-down electrons in the 8$\times$2 quarter-partition, $A$, of our system. These probabilities reflect the fact that, while the filling of the entire system remains fixed, the number of spin-up and spin-down electrons in the partition may fluctuate.  As we further illustrate below, different partitions may be selected to highlight different competing orders (see Fig.~\ref{fig:heatmap_half_filling}(b)). However, we begin with the simplest contiguous partition which illustrates how electrons fluctuate in and out of a uniform area of the lattice. At sufficiently large interaction strengths $|U|$, the fluctuations in the population of spin-up and spin-down electrons in the partition are not random, but instead reflect the underlying strong correlations.

\begin{figure*}
  \includegraphics[width=0.75\textwidth]{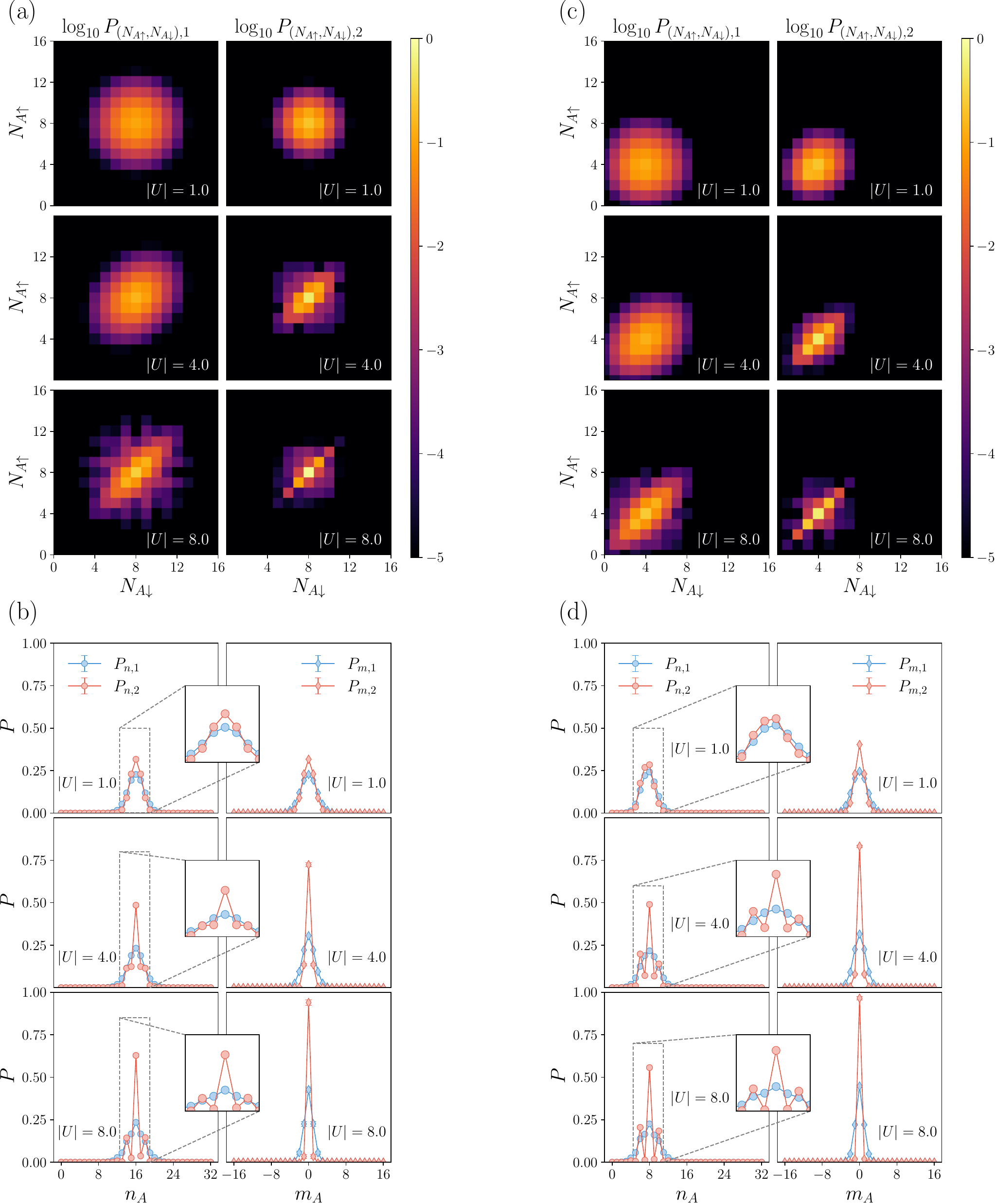}
  \caption{Various probability distribution measures plotted in an $8\times2$ quarter-partition of a $8\times 8$ Hubbard model at (a,c) half-filling and (b,d) quarter-filling with different $|U|$ values. (a-b) Joint (unresolved) probability distributions as a function of the spin-up ($N_{A\uparrow}$) and spin-down ($N_{A\downarrow}$) electron numbers. Left panel: $P_{(N_{A\uparrow},N_{A\downarrow}),1}$, the $\alpha=1$ distribution. Right panel: $P_{(N_{A\uparrow},N_{A\downarrow}),2}$, the $\alpha=2$ distribution. The color bar represents the range of probabilities on a log scale. (c-d) Charge-resolved (left panel) and spin-resolved (right panel) probabilities distributions as a function of local charge ($n_A = N_{A\uparrow} + N_{A\downarrow}$) and spin number ($m_A = N_{A\uparrow} - N_{A\downarrow}$). Both the left panel and its insets demonstrate the enhanced pairing effect reflected in fluctuations for even and odd charge numbers as $|U|$ increases.}
  \label{heatmaps}
\end{figure*}
 
In Fig.~\ref{heatmaps}\BR{(a)}, we present heatmaps of the logarithm of the joint probability distributions as a function of the number of spin-up and spin-down electrons in a quarter partition for a range of interaction strengths: $|U|=1$, $|U|=4$, and $|U|=8$. All of the distributions are centered around $N_{A\uparrow}=N_{A\downarrow}=N/8$, which stems from the fact that, for a quarter-partition and assuming translational symmetry, one expects an average of $1/4$ of the total $N/2$ electrons of each flavor to occupy the partition. On the left of the figure, we present the joint probability distribution functions for \ren index $\alpha=1$, while the right shows those for $\alpha=2$. The $\alpha=2$ cases are significantly more peaked and anisotropic than the corresponding $\alpha=1$ distributions. This remains true even in the absence of interactions where the $\alpha=2$ distributions are proportional to the square of the $\alpha=1$ distributions. The $\alpha=2$ distributions thus show enhanced resolution, regardless of $|U|$ \cite{Barghathi:2018rg}.
 
The increased sharpness of these $\alpha=2$ distributions makes them more sensitive to correlations. As can be seen for attractive interactions as small as $|U|=4$,  the $\alpha=2$ plots show larger probability amplitudes along the diagonal from small particle numbers (small $N_{A\downarrow}$, $N_{A\uparrow}$) to large particle numbers (large $N_{A\downarrow}$, $N_{A\uparrow}$). This diagonal character is a manifestation of the electron pairing expected in the attractive Hubbard model; as the number of spin-up electrons in the partition increases, so does the number of spin-down electrons because they are more likely to reside in the partition as pairs. As one might anticipate, the diagonal character of the distribution increases significantly with $|U|$, becoming almost completely diagonal for $|U|=8$ and indicating an increased preference for pairing. Indeed, while all of the distributions reach their maxima at $N_{A\uparrow}$=$N_{A\downarrow}=N/8$, we see that with increasing $|U|$, this maximum becomes increasingly pronounced, indicating that pairing favors a net magnetization of $0$. While this trend toward pairing is evident in both the $\alpha=1$ and $\alpha=2$ probability distributions, its clearer manifestation in the $\alpha=2$ distributions demonstrates their greater ability to infer subtle electron-electron interactions.

Looking more closely at the $|U|=4$ and $|U|=8$ distributions, it can be observed that they are non-monotonic for fixed $N_{A\downarrow}$ and varying $N_{A\uparrow}$ (or vice-versa).  For example, $N_{A\downarrow}=10$ is more likely to be accompanied by $N_{A\uparrow}=8$ or $N_{A\uparrow}=10$ than $N_{A\uparrow}=9$, resulting in even/odd oscillations in the probability distribution. This illustrates a preference toward integer magnetization: pairs of electrons, even if they do not possess opposite spins, are more likely to reside in the partition than unpaired electrons.

Although obtaining these joint distributions represents an algorithmic achievement and they reveal fascinating, multidimensional correlations, it is often easier to analyze their marginalized, one-dimensional versions.  Marginalization also grants us access to the individually charge- and spin-resolved quantities by summing over all possible total spin (magnetization) values given a fixed charge (particle number) and all possible charge values given a fixed total spin. In Fig.~\ref{heatmaps}(b), we show the charge- ($P_{n}$) and spin-resolved ($P_{m}$) distributions on the left and right, respectively. For each $|U|$ value, we plot both the $\alpha=1$ and $\alpha=2$ marginalized distributions to further highlight their differences. 

From these plots, it is even more evident that $P_{n,2}$, the charge-resolved distribution for $\alpha=2$, and $P_{m,2}$, the spin-resolved distribution for $\alpha=2$, are more peaked than their $\alpha=1$ counterparts. Also evident is the fact that the spin-resolved distributions tend to be more highly peaked than the charge-resolved distributions; the model overwhelmingly prefers a net magnetization of $0$, in concert with the highly diagonal joint distributions described earlier. In general, the charge distributions tend to be wider, suggesting that the model, which does not possess any direct charge-charge interactions, admits larger charge fluctuations than spin fluctuations. Nonetheless, as $|U|$ grows, even the charge distributions become increasingly narrow, pointing to higher-order terms that effectively cause the tightly-bound electron pairs to repel one another. Indeed, in the limit of infinite $|U|$, one expects the model to be equivalent to a hard-core boson model \cite{Ho:2009}, in which the paired spins act like bosons that effectively repel one another because they cannot occupy the same sites at the same time. In the absence of any explicit intersite interactions in Eq.~\eqref{eq:AHM}, the tightly-bound bosons can freely fluctuate between empty neighboring sites. 

The $P_{n,2}$ distributions also demonstrate non-monotonic behavior for strong interactions as seen in the insets of Fig.~\ref{heatmaps}(b) where oscillations around the most probable occupancy of $n_A=16$ are visible at $|U|=4$. These oscillations show that odd numbers of particles -- which would necessarily have to be unpaired -- are highly unlikely to be found in the partition. While the $P_{n,1}$ distributions do begin to show hints of non-monotonicity at $|U|=8$, the fact that these features are far more prominent in the $P_{n,2}$ distribution further demonstrates the value of this new measure.

To probe their sensitivity to filling fraction, in Figs.~\ref{heatmaps}(b) and (d), we present the same joint and marginalized distributions for a quarter-filled attractive Hubbard with the same contiguous quarter partition. These distributions bear many of the same features as seen in panels (a) and (c) at half-filling, despite the fact that the quarter-filled model is not expected to exhibit any charge density wave ordering. The joint distributions possess similar widths and maxima, but are centered around $N_{A\uparrow}=N_{A\downarrow}=4$ because only half as many electrons reside in the partition on average. The trend that the $\alpha=2$ case has sharper features is also maintained, and the marginalized distributions manifest oscillations at similar $|U|$ values. This suggests that, while the contiguous quarter partition can provide insights into the pairing present in the model, it cannot discern potential charge density wave ordering. We will discuss a different partition specifically designed to identify this ordering in Section \ref{checkerboard_sec}. 

\subsection{Symmetry-Resolved and Accessible Entanglement Entropies for a Quarter-Partition} 

\begin{figure*}
  \includegraphics[width=0.9\textwidth]{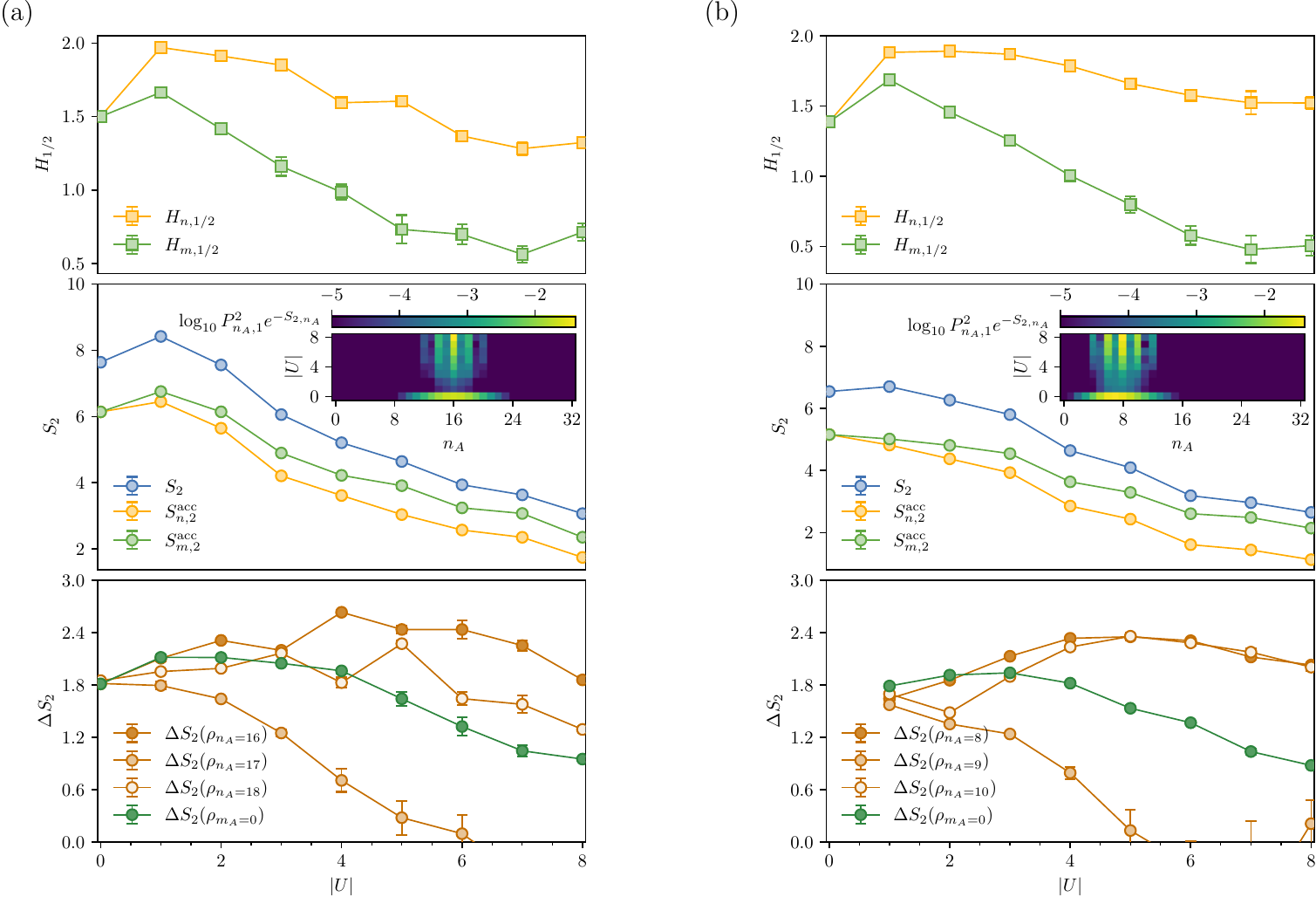}
  \caption{Entanglement measures plotted at (a) half-filling and (b) quarter-filling against interaction strength $|U|$ for an $8\times8$ lattice with a contiguous $8\times2$ partition. Upper panel: Generalized Shannon entropies resolved according to both the charge and spin. Middle panel: R{\'e}nyi-2 entropy plotted alongside the charge- and spin-resolved accessible entanglement entropies. Lower panel: Symmetry-resolved entanglement entropy for three distinct partition occupancies (charges): (a) $n_A=16,17,18$ and (b) $n_A=8,9,10$ and the magnetization at $m_A=0$. The inset heatmap depicts the contribution from different charge sectors to the overall R{\'e}nyi-2 entropy, $P_{n_A,1}^2e^{-S_{2,n_A}}$, on a log scale.}
  \label{fig:entangle}
\end{figure*}

With these charge- and spin-resolved distributions in hand, we can now leverage Eqs.~\eqref{S_alpha}, \eqref{eqn:AccRenyiDecomposition}, and \eqref{eqn:Shannon} to compute the Shannon and accessible entanglement entropies. As described above, the generalized Shannon entropies, $H_{1/\alpha}$, can be viewed as measures of the entropy stemming from local fluctuations of a globally conserved quantity in and out of the partition. In the top panel of Fig. \ref{fig:entangle}(a), we plot the charge- and spin-resolved $H_{n,1/2}$ and $H_{m,1/2}$ entropies as a function of $|U|$ for the half-filled attractive Hubbard model on a contiguous quarter partition. In line with the previously discussed probability distributions, we see that the charge-resolved Shannon entropy is consistently larger than the spin-resolved Shannon entropy. This means that the particle number fluctuates more readily than the spin, consistent with the narrow peak seen in $P_{m,1}$ (Fig.~\ref{heatmaps}(b)). Both entropies are also observed to generally increase as $U$ approaches $0$, the non-interacting limit in which the particles have the most freedom to fluctuate and the least correlation. At larger values of $|U|$, the electrons have much less freedom to unpair and their contributions from fluctuations to the entropy are reduced. 

Based on Eq.~\eqref{eqn:AccRenyiDecomposition}, the \ren entanglement entropy and the Shannon entropies grant us access to the \ren accessible and symmetry-resolved entanglement entropies.  These are plotted in the middle panel of Fig.~\ref{fig:entangle}(a) along with the unresolved entanglement, which is largest as expected due to the additional constraints these symmetry-resolved entanglements reflect. For small interaction strengths ($0\leq\vert U\vert\leq1.0$), all entanglement measures slightly increase with $\vert U\vert$, but beyond $\vert U\vert\approx1.0$ all entanglement measures monotonically decrease with increasing $|U|$, indicating that the electrons are most entangled at smaller $|U|$ values. The increased pairing at larger $|U|$ suppresses the available degrees of freedom resulting in decreased entanglement overall. In Fig.~\ref{fig:entangle}(a), it can furthermore be seen that the charge- and spin-resolved accessible entanglements follow roughly the same trends, but are smaller overall because they report on only the accessible contributions (\emph{i.e.}, the entanglement solely due to fluctuations is removed). Because the charge-resolved Shannon entropy was previously seen to be larger than the spin-resolved entropy, the charge-resolved accessible entanglement ends up being smaller than the spin-resolved accessible entanglement. 

In the bottom panel of Fig.~\ref{fig:entangle}, we show a detailed comparison of the particle number-resolved and total entanglement entropies by considering the difference between them: $\Delta S_{2}$, as defined by Eq.~\ref{eq:entang_diff}. 
We consider the particle number-resolved entanglements for specific particle numbers in the partition, $n_{A}$, with the aim of identifying which particle numbers contribute most to the entanglement. Our ability to analyze the contributions to the entanglement entropy from different charge and spin sectors is a highlight of the algorithms presented here that affords highly detailed information regarding the underlying origin of the observed entanglement. For example, for small interaction strengths, all $\Delta S_{2}$ values are similar, but as $|U|$ increases, clear differences emerge. Of the particle number-resolved plots in orange, the $\Delta S_{2}(\rho_{n_{A}=16})$ and $\Delta S_{2}(\rho_{n_{A}=18})$ values remain the largest irrespective of $|U|$ because the entanglement for those particle numbers is smallest since the electrons are strongly paired at those occupations. In contrast, $\Delta S_{2}(\rho_{n_{A}=17})$ rapidly decreases as $|U|$ increases because the entanglement in that sector remains large due to unfavorable occupancies resulting in particle fluctuations. $\Delta S_{2}(\rho_{m_{A}=0})$ also slightly decreases with increasing $|U|$ due to the reduced width of the spin-resolved distribution, as observed in Fig.~\ref{heatmaps}(d). However, the preference toward pairing prevents fluctuations in the net spin, resulting in $\Delta S_{2}(\rho_{m_{A}=0})$ remaining sizeable, even at $|U|=8$.

To more thoroughly ground this discussion in the formalism presented in Sec.~\ref{formal}, recall that the particle number-resolved reduced density matrix, $\rho_{n_A}$, is still subject to magnetization fluctuations and can be further resolved as 
\begin{eqnarray}
    \rho_{n_A} &=& \sum_{m_A} \hat{\Pi}_{A_m}\rho_{n_A}\hat{\Pi}_{A_m} \nonumber\\
    &=& \sum_{m_A}P(m_A \vert n_A)\rho_{n_A,{m_A}}, 
\end{eqnarray}
where $P(m_A \vert n_A)$ is the corresponding conditional probability and $n_A$ and $m_A$ must have the same parity. This suggests that, in the strong pairing regime, the magnetization resolution of the RDMs $\rho_{n_A=16}$ and $\rho_{n_A=18}$ is dominated by the RDMs $\rho_{n_A=16,m_A=0}$ and $\rho_{n_A=18,m_A=0}$, respectively. Thus,  the contribution from the magnetization fluctuations to the corresponding entanglement is suppressed. In contrast, the magnetization resolution of RDM $\rho_{n_A=17}$ contains contributions from odd $m_A$ only, but with equal contributions from the positive and negative $m_A$ sectors. If only the contributions from $m_A=\pm1$ are considered, then the resulting entanglement will be $\ln2$ greater than the entanglement content of $\rho_{n_A=17,m_A=1}$. In addition, because of the strong pairing between the electrons, the RDM $\rho_{n_A=17,m_A=1}$ is expected to have more entanglement than $\rho_{n_A=18,m_A=0}$ due to the assured presence of at least one correlated broken pair across the partition boundary. The above discussion indicates that strong pairing between the electrons could produce oscillations in the symmetry-resolved entanglement due to changes in the parity of the particle number. 

Perhaps most telling is the inset of the middle panel of Fig. \ref{fig:entangle}(a) in which we plot the different contributions to the trace of $\rho_A^2$ on a log scale,
\begin{equation}
    \Tr\rho_A^2 = e^{-S_2} = \sum_{n_A} P^2_{n_A, 1}e^{-S_{2,n_A}},
\end{equation}
which reflects a combination of the $P_{n,1}^{2}$ and $S_{2,n_A}$ components. As is evident from the heat maps, at $\vert U\vert=0$, the contributions from different $n_A$ sectors are slowly varying. By increasing $\vert U\vert$, we observe a peaked distribution at $n_A = 16$ which begins to develop oscillations by $\vert U\vert=4$. However, as we have seen in Fig.~\ref{heatmaps}(b), $P_{n,1}$ does not develop oscillations even at $\vert U\vert=8$, which strongly indicates that the sources of the observed oscillations are the symmetry-resolved entanglements, $S_{2,n_A}$. 

Similar behavior is again seen in the corresponding quarter-filled AHM shown in Fig.~\ref{fig:entangle}(b), indicating that many of the same mechanisms that give rise to the half-filled physics are also at play at quarter-filling.  One noticeable difference is that the discrepancy between $\Delta S_{2}(\rho_{n_{A}=16})$ and $\Delta S_{2}(\rho_{n_{A}=18})$ in the half-filled case is significantly larger than that between $\Delta S_{2}(\rho_{n_{A}=8})$ and $\Delta S_{2}(\rho_{n_{A}=10})$ in the quarter-filled case, where $\Delta S_{2}(\rho_{n_{A}=8})$ and $\Delta S_{2}(\rho_{n_{A}=10})$ are almost equal. We speculate that this is a signature of the influence of the charge density wave ordering in the half-filled case. Based on a previous argument, the main contributions to $S_{2}(\rho_{n_{A}=16})$  and $S_{2}(\rho_{n_{A}=18})$ are from $\rho_{n_A=16,m_A=0}$ and $\rho_{n_A=18,m_A=0}$, respectively. While the RDM $\rho_{n_A=16,m_A=0}$ is favored by charge density wave ordering ($\rho_{n_A=16,m_A=0}$ describes 8 pairs of electrons in the quarter partition), the RDM $\rho_{n_A=18,m_A=0}$ violates such ordering by including an extra pair of electrons, which is expected to increased $ S_{2}(\rho_{n_{A}=18})$. While this provides a possible justification for the observed differences in the plots of the symmetry-resolved entanglements, it can not be considered definitive evidence for the presence of charge density wave ordering at half-filling. This suggests that other partitions may be necessary to unambiguously identify the charge density wave ordering expected to be present at half-filling.   

\subsection{Probability Distribution Functions for a Checkerboard Partition \label{checkerboard_sec}}  

In the previous sections, we presented results for the generalized entanglement entropies of the attractive Hubbard model on a partition defined by a contiguous block of a quarter of the lattice sites. One of the benefits of studying entanglement entropies is that they can be calculated for different partitions which can be engineered to probe different underlying physical phenomena. At half-filling, the electrons in the attractive Hubbard model experience a competition between pairing and charge density wave ordering in the ground state. To study the emergence of charge density wave order, which was demonstrated to be difficult to discern based on the quarter-filled partitions analyzed above, in the following we examine the same probability distribution functions and entanglement entropies, but on a \emph{checkerboard partition} that includes every other site (see Fig.~\ref{fig:heatmap_half_filling}): if a pure checkerboard ordering emerges, all of the electrons will either occupy the partition or its complement (due to translational symmetry). Because charge density wave order should be less favorable at lower fillings, one would not expect the checkerboard partition to be particularly informative at quarter filling. The checkerboard partition should hence be able to distinguish between half-filled and quarter-filled physics.   

\begin{figure}
    \centering
    \includegraphics[width=0.4\textwidth]{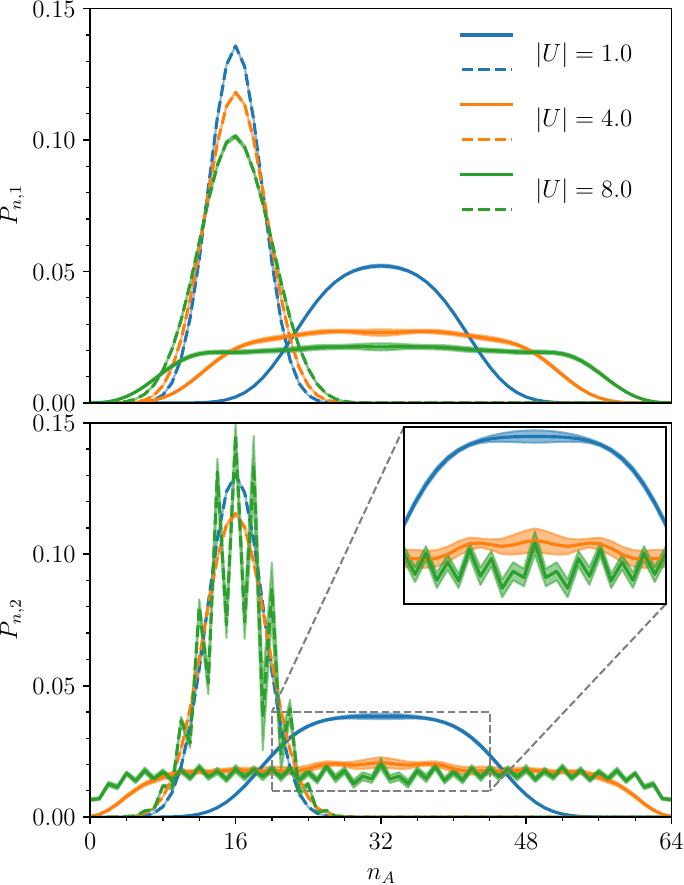}
     \caption{Charge-resolved probabilities distribution functions ($P_{n,1}$ and $P_{n,2}$) as a function of local charge for an $8\times 8$ Hubbard model at half-filling (solid lines) and quarter-filling (dashed lines) measured using the checkerboard bipartition. For clarity, the error bars are represented as ribbons around the curves.}
    \label{checkerboard_fig}
\end{figure}

In Fig.~\ref{checkerboard_fig}, we plot the particle number-resolved probability distribution functions for the checkerboard partition for different interaction strengths: $\abs{U}=1.0$, $4.0$, and $8.0$. In the top panel, we show the $P_{n,1}$ distributions at quarter- (dashed curves) and half-filling (solid curves). At quarter-filling, the distributions remain roughly Gaussian around $n_A=16$ for all $|U|$ values. The $\abs{U}=1$ distribution is the most peaked with the distributions broadening with decreasing $U$ because of the buildup of positive correlations among the electrons as they start to form pairs. The absence of odd $n_A$ versus even $n_A$ oscillations, as were observed in the contiguous partition, can be attributed to the much larger boundary of the checkerboard partition: four sites of the complementary partition surround each site in the checkerboard partition. Thus, any broken electron pair at any site will result in an odd $n_A$ contribution to the distributions, reducing the magnitude of the oscillations observed. Taken together, these features signify that no particular ordering emerges within the checkerboard partition at quarter-filling. 

In contrast, the distributions at half-filling exhibit a flattening with increasing interaction strength. These broad distributions are indicative of the emergence of charge density waves. 
\begin{figure}[ht]
    \centering
    \begin{subfigure}[b]{0.45\textwidth}
        \centering
        \includegraphics[width=\textwidth]{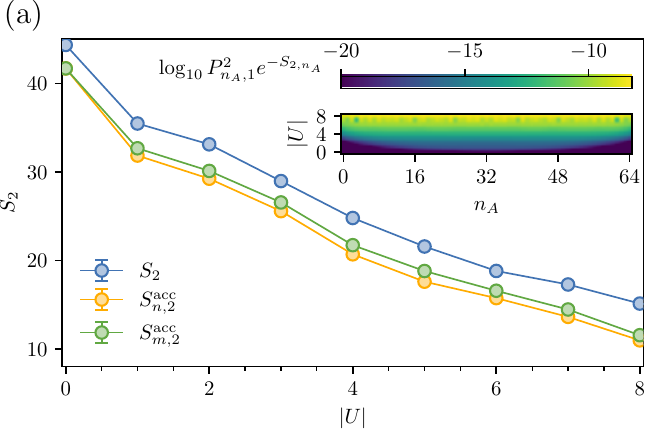}
    \end{subfigure}
    \begin{subfigure}[b]{0.45\textwidth}
        \centering
        \includegraphics[width=\textwidth]{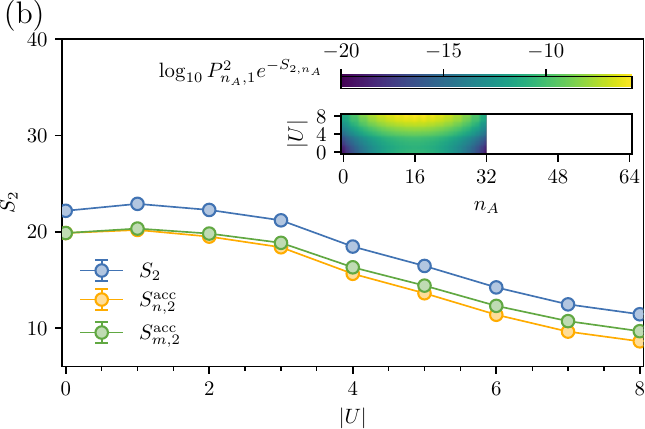}
    \end{subfigure}
    \caption{R{\'e}nyi-2 charge- and spin-resolved accessible entanglement entropies as a function of $|U|$ for an $8\times 8$ attractive Hubbard model at (a) half-filling and (b) quarter-filling with a checkerboard partition. For clarity, The inset heatmap depicts the log scale contribution from different charge sectors to the overall R{\'e}nyi-2 entropy, $P_{n_A,1}^2e^{-S_{2,n_A}}$. For the heatmap in (b), the right half is blank as the local electron number can only reach a maximum of 32 at quarter filling.}
    \label{checkerboard_ent}
\end{figure}
In particular, at $\abs{U}=1$, at which minimal charge ordering would be expected to occur, most of the probability density is located between $n_A=16$ and $n_A=48$, while at $\abs{U}=8$, well into the regime in which charge ordering should manifest, the probability density is distributed between $n_A=2$ and $n_A=62$. The $P_{n,1}$ distributions presented in the top panel of Fig.~\ref{checkerboard_fig} overwhelmingly appear smooth. However, as shown in the bottom panel, the high resolution provided by $P_{n,2}$ shows that these distributions do in fact possess fine features - namely the even-odd oscillations discussed in Section~\ref{jointmargin} and even finer features unique to the checkerboard partition. Interestingly, different even occupations, even for moderate $n_{A}$ values, are observed more or less frequently than others, suggesting that not only is pairing preferred, but very specific types of pairing are favored that maintain the emerging charge density wave ordering. These $P_{n,2}$ distributions can thus uniquely capture competition between two forms of order in a single measure. 

As shown in Fig.~\ref{checkerboard_ent}, differences reflected by the use of a checkerboard partition may also be observed in the entanglement entropies. $S_{2}$ as a function of $|U|$ decays in a similar manner as in the contiguous quarter partition, but starts at larger values at $|U|=0$ owing to the checkerboard partition's larger area. More tellingly, the contributions to the entanglement entropy resolved by the charge in the partition assume a different form than observed with the previous partition. As the heatmap in the inset depicts, the contribution to the entanglement is smallest for small $|U|$ and increases for large $|U|$. As discussed before, this reflects a balance between contributions from $S_{2}$ and $P_{n,1}^{2}$. In this case, the entanglement is large for small $|U|$, which decreases the magnitude of the contributions to the heatmap in the weak interaction limit. At larger $|U|$, the entanglement decreases, increasing the contributions to the heatmap. The larger intensity region of the heatmap also broadens at larger $|U|$ due to the wider $P_{n,1}$ distributions observed at these values. 

Overall, the results on the checkerboard partition illustrate the utility of being able to define a variety of partitions that can report on different facets of the system's propensity towards cooperative order. In this case, the checkerboard partition highlights differences that could not be fully observed using a contiguous partition, or even, certain correlation functions, as will be discussed below. 

\section{Comparing the Information Content in Symmetry-Resolved Entanglement Entropies and Correlation Functions}

\begin{figure}
\label{corrfxns}
  \centering
  \begin{minipage}{0.5\textwidth}
    \includegraphics[width=0.78\linewidth]{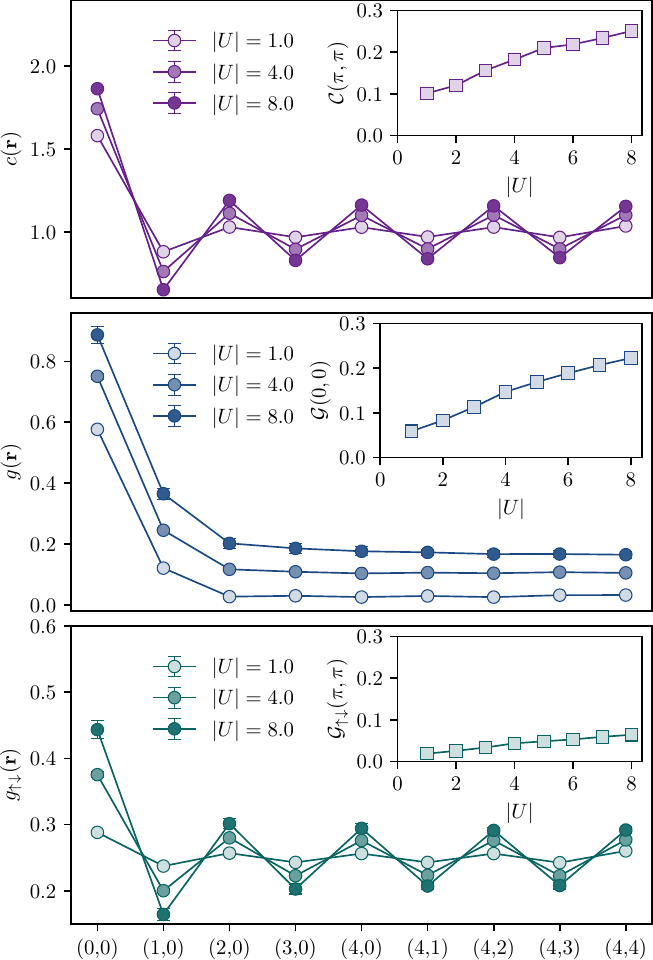}
    \caption{Correlation functions and structure factors for an $8 \times 8$ attractive Hubbard model at half-filling. The charge-charge, pair, and spin-resolved pair correlation functions are measured along a triangle path in the lattice. The insets depict the corresponding structure factors as a function of interaction strength $|U|$ measured at ${\bf k} = (\pi,\pi)$, $(0,0)$, and $(\pi,\pi)$, respectively.}
    \label{fig:corr_funcs_half_filling}
    \vspace{1.5em}
    \includegraphics[width=0.78\linewidth]{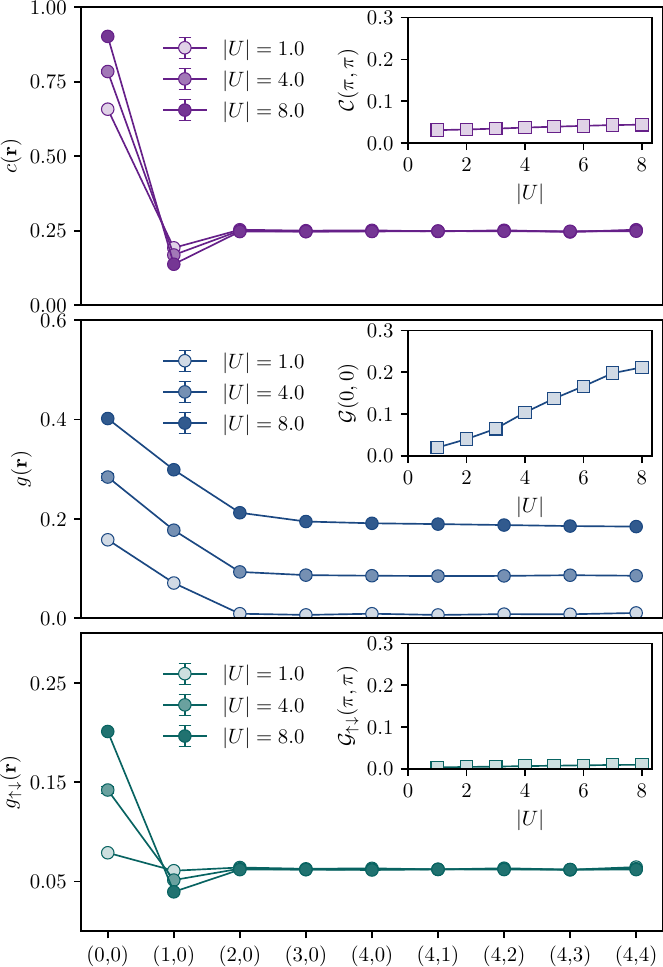}
    \caption{Correlation functions and related structure factors for the quarter-filled attractive Hubbard model.}
    \label{fig:corr_funcs_quarter_filling}
  \end{minipage}
\end{figure}

Given the complexity and computational expense that accompanies computing entanglement entropies, a natural question is whether they have the potential to add anything new to our understanding beyond what can be learned from conventional local correlation functions. In particular, for the attractive Hubbard model that exhibits pairing, and at half-filling, a charge density wave, we seek to understand whether entanglement entropies on designer partitions can signal larger and/or more rapidly developing features upon the emergence of different types of ordering.

To make this comparison, we analyze three local correlation functions and their corresponding structure factors. These include the charge-charge correlation function
\begin{equation}
    C({\bf r}) = \frac{1}{N_s} \sum_{\bf i} \expval{n_{\bf i + r} n_{\bf i}},
\end{equation}
the pair correlation function
\begin{equation}
    g({\bf r}) = \frac{1}{N_s} \sum_{\bf i} \expval{c^\dagger_{{\bf i+r}, \uparrow} c^\dagger_{{\bf i}, \downarrow} c_{{\bf i+r}, \downarrow} c_{{\bf i}, \uparrow} + \rm{H.c.}},
    \label{pair}
\end{equation}
and the spin-resolved pair correlation function
\begin{equation}
    g_{\uparrow\downarrow}({\bf r}) = \frac{1}{N_s} \sum_{\bf i} \expval{n_{{\bf i+r}, \uparrow} n_{{\bf i}, \downarrow}}.
\end{equation}
Their Fourier transforms (structure factors) are denoted as $\mathcal{C}(\bf k)$, $\mathcal{G}(\bf k)$, and $\mathcal{G}_{\uparrow\downarrow}(\bf k)$, respectively. As shown in Fig.~\ref{fig:corr_funcs_half_filling}, the correlation functions at half-filling capture the emergence of charge ordering in the system in the form of oscillations. The magnitude of these oscillations grows with $|U|$, as is also reflected in the associated structure factors at wavevector $(\pi,\pi)$ plotted in the insets.  These oscillations are additionally observed in the spin-resolved pair correlation function (bottom panel), which indicates that spins also alternate from site to site as part of the pairs formed,  although with a smaller magnitude than the charges. The pair correlation function provided in Eq.~\eqref{pair} (middle panel) measures the correlation of pairs of spin-up and spin-down electrons at increasing separations between the paired fermions. As one would anticipate, these pairs are most correlated at zero separation and for larger $|U|$ values. The correlation decays with increasing distances between the electrons in the pairs, but seemingly plateaus to finite values that grow with $|U|$. These plateaus suggest that long-range order is stabilized and the pairs of electrons begin to approximate hard-core bosons at large $|U|$. The Fourier transforms of all of the order parameters presented in the insets all correspondingly grow with $|U|$. 

In Fig.~\ref{fig:corr_funcs_quarter_filling}, the same correlation functions and structure factors are plotted, but for the AHM at quarter-filling. These correlation functions do not manifest persistent oscillations, substantiating the expectation that charge ordering does not emerge at quarter-filling.

Altogether, these plots confirm the emergence of charge density waves and pairing only at half-filling, as also observed in the entanglement entropies described above -- albeit based on very different signatures. The correlation functions and structure factors provide important information about how the electrons arrange themselves in real and reciprocal space. In contrast, the symmetry-resolved entanglements computed above provide information about how certain collections of particles (e.g., with fixed numbers or magnetizations) behave in concert and thereby are most entangled. Within our formalism, spatial information can only be gleaned from the entanglements if it is reflected in the partition. Thus, the correlation functions and symmetry-resolved entanglements lie on two ends of the information spectrum: correlation functions average over all types of configurations, but provide useful spatial information, while the symmetry-resolved entanglement entropies average over spatial information, but provide useful configurational information. These measures hence provide complementary perspectives and choices can be made as to which to use based upon the features to be highlighted.

\section{Conclusions}
\label{conc}

In this manuscript, we have introduced a numerical method based on Auxiliary Field Quantum Monte Carlo to compute the accessible and symmetry-resolved entanglement entropies of interacting systems of fermions, as well as their associated joint and marginalized probability distributions that capture fluctuations of local degrees of freedom subject to a global constraint.  We apply this method to the two-dimensional attractive Hubbard model in the presence of fixed total particle number and magnetization to probe how quantum information measures may capture signatures of pairing and emergent charge density wave order in the ground state of a strongly interacting fermionic system.  Key to being able to apply these methods to the attractive Hubbard model was the use of a recursive algorithm that enables the calculation of observables at fixed particle numbers \cite{Shen_PRE_2023} and an incremental swap algorithm for computing the R{\'e}nyi-2 entanglement entropy with improved statistical convergence \cite{da2023controllable}.  Using the flexibility provided by our simulation method, we studied different spatial bipartitions of the two-dimensional square lattice.  We found that, while a contiguous quarter partition was useful for seeing evidence of pairing, a checkerboard partition was important for seeing evidence of charge density waves. The ability to compute entanglements symmetry-resolved into different charge and spin sectors allowed for novel insights into which superselection sectors contribute most to the entanglement. Interestingly, the $\alpha=2$ probability distributions and entanglement entropies showed enhanced detail, exhibiting greater signatures of the underlying correlations and ordering, than their $\alpha=1$ counterparts. This is important as higher order \ren entropies can be readily computed using quantum Monte Carlo methods (as well as in cold atom experiments \cite{Islam:2015ap,Pichler:2016rd, Lukin:2019dy, Brydges:2019zv}), in contrast with the von Neumann entropies that often require full state tomography. 

Overall, the symmetry-resolved and related entanglement measures studied here provide a complementary picture to that exhibited by traditional correlation functions.  For example, while conventional correlation functions provide information about relatively local physics and ordering, entanglement measures provide access to non-local information about collections of particles with conserved total charge and magnetization.  The ability to create custom partitions of the many-body quantum state in combination with the projection of entanglement into different symmetry sectors provides direct access to the relative importance of fluctuations vs. configurations of the relevant degrees of freedom. This valuable information sheds a bright light on the underlying energy-entropy competition responsible for non-classical ordering phenomena.  

While the formalism presented here provides new means of measuring entanglement in the study of many-body systems (including, potentially, in experiments on ultracold lattice gasses), many questions about its applications remain. Firstly, we showed how it can be crucial to define different types of partitions with different sizes and geometries for interrogating various types of physical phenomena (e.g., charge ordering and pairing).  This is analogous to constructing appropriate correlation functions that are sensitive to different types of local ordering.  How these partitions can be engineered to best and most efficiently illuminate different types of ordering remains to be explored. For example, by designing ensembles of two-site partitions, it may be possible to construct an ``entanglement matrix'' analogous to conventional correlation matrices consisting of the two-site correlation functions between all pairs of sites. Such a matrix would provide valuable insights into the distance scaling of different forms of entanglement. 

In this work, we focused on a paradigmatic model displaying coexistent ordering in its ground state, but lacking any topological character. Given that the entanglement is a more global measure of electronic behavior, it is likely that the accessible and symmetry-resolved entanglements studied here would provide even greater insights into order emerging near phase transitions that induce long length-scale fluctuations and topological order invisible to local correlation functions. Indeed, one could imagine using these entanglement measures to probe the subtle onset of ordering that may appear in only select charge or spin sectors upon approaching different or multiple phase boundaries.  The fact that the particle number distributions on the checkerboard partition are simultaneously sensitive to both charge density wave order and pairing is a signal that these entanglement measures could potentially probe and resolve the competition among multiple emergent orders in the same measure. This would be particularly useful for analyzing models and materials that exhibit both strong electron correlation and topology, such as correlated topological insulators \cite{Mai:2023ox}.

\section{Data Availability}

All codes, scripts, and data needed to reproduce the results in this paper are available online \cite{source_code, papers_code_repo}. 

\section{Acknowledgements \label{ack}}
The authors thank Ben Cohen-Stead for insightful conversations. T.S., J.Y., and B.R. were funded by NSF CTMC CAREER Award 2046744 and A.D.\@ acknowledges support from the
U.S.\@ Department of Energy, Office of Science, Office of Basic Energy Sciences,under Award No. DE-SC0022311.  T.S.\@ is also grateful for financial support from the Brown Open Graduate Education program. This research was conducted using computational resources and services at the Center for Computation and Visualization, Brown University.

\appendix
\section*{Appendix A: Replica Sampling Statistics}
\label{replica}

\renewcommand\thefigure{A.\arabic{figure}}
\setcounter{figure}{0}

\begin{figure*}[t]
    \centering
    \includegraphics[width=0.8\textwidth]{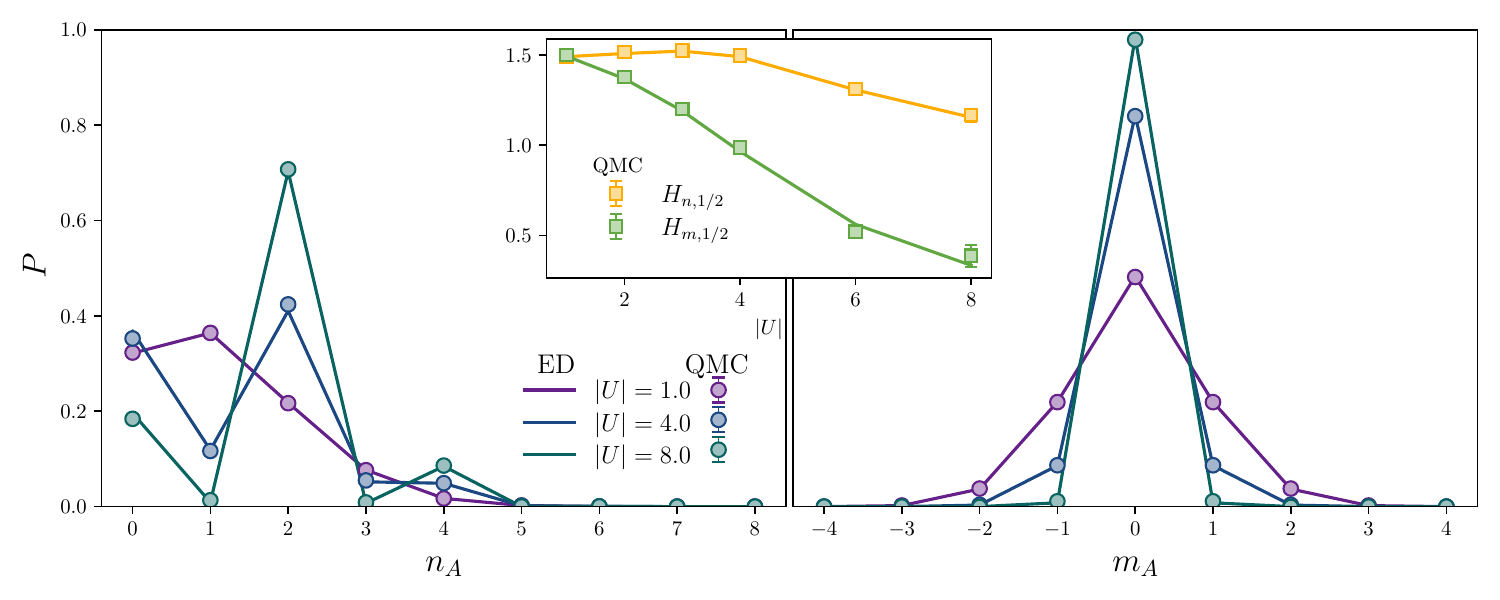}
     \caption{Comparison of the $P_{n,2}$ (left panel) and $P_{m,2}$ (right panel) distributions obtained using replica sampling and exact diagonalization (ED) for the $4\times3$ Hubbard model at quarter filling, in which the subsystem is chosen as a $4\times 1$ strip. The inset shows the estimated generalized Shannon entropy compared to ED for various $|U|$ values.}
    \label{ED_comparison_fig}
\end{figure*}

In this appendix, we demonstrate the numerical stability of our replica sampling technique for estimating $P_{q,2}$, as described in Eq.~\eqref{eqn:P_q2_recursion}. For validation, we compare QMC estimates of $P_{q,2}$ and the derived Shannon entropies ($H_{1/2}$) with exact results obtained from Exact Diagonalization (ED). This comparison is conducted on a $4\times3$ Hubbard model at quarter filling. To minimize Trotter errors, we employ a small Trotter step ($\Delta\tau = 0.01$) and set the projection time to $\Theta = 20$, which we find sufficient for ground state projection. The QMC data presented are averaged over 5,120 independent samples, with Shannon entropies estimated through the Jackknife resampling method.

Fig. \ref{ED_comparison_fig} reveals near-perfect agreement between QMC and ED for $P_{n,2}$ and $P_{m,2}$ within QMC error bars. However, a small discrepancy is observed in the spin-resolved Shannon entropy, $H_{m,1/2}$, at $|U|=8$. This discrepancy arises because, in the strong $|U|$ regime, only the $m_A=0$ sector contributes to $H_{m,1/2}$ due to the strong electron pairing. As $P_{m_A=0,2}$ asymptotically approaches 1, the precision with which QMC can reproduce this value diminishes due to error propagation, leading to an observable deviation of the estimated $H_{m,1/2}$ from the exact result. Furthermore, reducing the error bar through replica sampling is inherently more challenging than plain sampling. This is due to the fact that the computation of the statistical weights for the field-dependent RDM, ${Z^{(2)}_A({{\bf s}_1}, {{\bf s}_2})}$, necessitates the sampling of two independent auxiliary fields. The fluctuations in these samples cover the space of two joint auxiliary fields, resulting in a probability distribution with a heavier tail, which in turn leads to an increased error bar.

\section*{Appendix B: Comparison of Incremental and Swap Algorithm Statistics}
\label{swap_and_incremental}
\numberwithin{equation}{section}
\renewcommand{\theequation}{B.\arabic{equation}}
\setcounter{equation}{0}
\renewcommand\thefigure{B.\arabic{figure}}
\setcounter{figure}{0}

\begin{figure}
    \centering
    \includegraphics[width=0.45\textwidth]{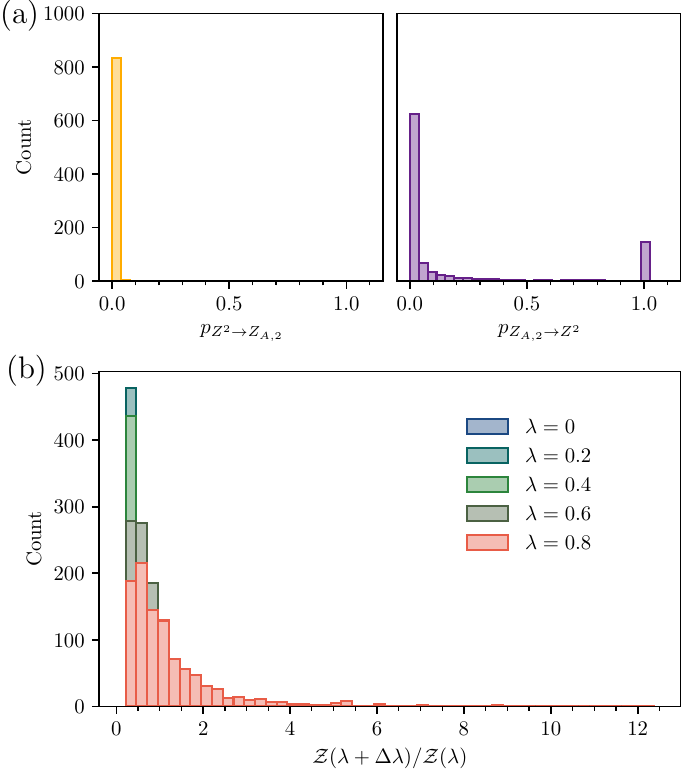}
     \caption{Histograms of 1024 uncorrelated samples for estimating $S_2$ using (a) the swap algorithm and (b) the incremental algorithm. In the swap algorithm, $S_2$ is estimated as $\frac{\expval{p_{Z^2\rightarrow Z_{A,2}}}}{\expval{p_{Z_{A,2}\rightarrow Z^2}}}$, while in the incremental algorithm, it is estimated as $\prod \expval{\mathcal{Z}(\lambda+\Delta\lambda) / \mathcal{Z}(\lambda)}$.}
    \label{Algo_comparison_fig}
\end{figure}

In this appendix, we conduct a comparative analysis of the sampling efficiency between two algorithms that enhance Grover's initial method for estimating \ren entropies: the incremental algorithm, as used in the main text and detailed in Eqs.~\eqref{eqn:incremental_def}, \eqref{eqn:incremental_prods}, and \eqref{eqn:incremental_estimator}, and the swap algorithm. To make this discussion self-contained, we first present an outline of the swap algorithm, and more detailed information can be found in Refs.~\cite{hastings2010measuring} and \cite{broecker2014renyi}. Similar to the incremental algorithm, the swap algorithm first rewrites $e^{-S_2(\rho_A)}$ as a ratio of partition functions
\begin{equation}
    e^{-S_2(\rho_A)} = \frac{\int \mathcal{D}{{\bf s}_1} \mathcal{D}{{\bf s}_2} Z_{A,2}({{\bf s}_1, s_2})}{\int \mathcal{D}{{\bf s}_1} \mathcal{D}{{\bf s}_2} Z^2({{\bf s}_1, {\bf s}_2})}, \label{eqn:SwapEstimator_1}
\end{equation}
with the shorthand $Z_{A,2}({{\bf s}_1, {\bf s}_2}) = Z_{{\bf s}_1} Z_{{\bf s}_2} \det g_A^{{{\bf s}_1}, {{\bf s}_2}}$ and $Z^2({{\bf s}_1, s_2}) = Z_{{\bf s}_1} Z_{{\bf s}_2}$. As the fluctuations in Grover's method comes from the unboundedness of the determinant of the Grover matrix, $\det g_A^{{{\bf s}_1}, {{\bf s}_2}}$, the swap algorithm re-expresses Eq.~\eqref{eqn:SwapEstimator_1} as the ratio of transition probabilities
\begin{equation}
    e^{-S_2(\rho_A)} = \frac{\expval{p_{Z^2\rightarrow Z_{A,2}}}_{Z^2}}{\expval{p_{Z_{A,2}\rightarrow Z^2}}_{Z_{A,2}}},
\end{equation}
where the transition probabilities are defined as
\begin{eqnarray}
    p_{Z^2\rightarrow Z_{A,2}}({{\bf s}_1, {\bf s}_2}) &=& \min(1, \frac{Z_{A,2}({{\bf s}_1, s_2})}{Z^2({{\bf s}_1, s_2})}), \\
    p_{Z_{A,2}\rightarrow Z^2}({{\bf s}_1, {\bf s}_2}) &=& \min(1, \frac{Z^2({{\bf s}_1, s_2})}{Z_{A,2}({{\bf s}_1, s_2})}),
\end{eqnarray}
and their expectation values are computed as
\begin{eqnarray}
    \expval{p_{Z^2\rightarrow Z_{A,2}}}_{Z^2} &=& \frac{\int \mathcal{D}{{\bf s}_1} \mathcal{D}{{\bf s}_2} Z^2({{\bf s}_1, s_2})p_{Z^2\rightarrow Z_{A,2}}({{\bf s}_1, {\bf s}_2})}{Z^2} , \nonumber \\
    \\
    \expval{p_{Z_{A,2}\rightarrow Z^2}}_{Z_{A,2}} &=& \frac{\int \mathcal{D}{{\bf s}_1} \mathcal{D}{{\bf s}_2} Z_{A,2}({{\bf s}_1, s_2})p_{Z_{A,2}\rightarrow Z^2}({{\bf s}_1, {\bf s}_2})}{Z_{A,2}}. \nonumber \\
\end{eqnarray}
The fluctuations from the unbounded $\det g_A^{{{\bf s}_1}, {{\bf s}_2}}$ are therefore reduced because the transition probabilities are bounded from above by $1$. In practice, one conducts two separate simulations that sample the ensemble of two independent replicas with weight $Z^2({{\bf s}_1, s_2})$ and  the ensemble of ``fused'' replicas with weight $Z_{A,2}({{\bf s}_1, s_2})$, respectively. The transition probabilities can be interpreted as the probabilities of swapping the random walk from one ensemble to another in the context of path integral Monte Carlo. \ren entropies can therefore be estimated from the ratio of transition probabilities.

In Fig.~\ref{Algo_comparison_fig}, we present histograms of $1024$ uncorrelated Monte Carlo samples obtained from both the swap and incremental algorithms to compare their sampling efficiencies. The same model as described in the main text is utilized, specifically an $8 \times 8$ attractive Hubbard model at half-filling, with a contiguous $8\times 2$ partition. The on-site interaction strength is fixed to $|U|=8.0$, a level sufficiently strong to induce significant fluctuations in the value of $\det g_A^{{{\bf s}_1}, {{\bf s}_2}}$. As observed in the left panel of Fig.~\ref{Algo_comparison_fig}(a), despite the extreme values of $\det g_A^{{{\bf s}_1}, {{\bf s}_2}}$ being capped at $1$ through probability mapping, the frequency of these extreme values is notably high at strong interaction strengths. This results in a heavy-tailed distribution. In contrast, using the incremental algorithm, the distribution tails decay more rapidly across all incremental values, $\lambda$. Consequently, the statistical errors are smaller in the incremental algorithm. Therefore, we have opted to use the incremental algorithm for all R{\'e}nyi-2 entropy calculations presented in the main text.

\bibliography{ref}{}
\end{document}